\documentclass[11pt]{article}

\newsavebox{\junk}
\savebox{\junk}[1.6mm]{\hbox{$|\!|\!|$}}

\def\bbbz{{\mathchoice {\hbox{$\sf\textstyle Z\kern-0.4em Z$}}
{\hbox{$\sf\textstyle Z\kern-0.4em Z$}}
{\hbox{$\sf\scriptstyle Z\kern-0.3em Z$}}
{\hbox{$\sf\scriptscriptstyle Z\kern-0.2em Z$}}}}
\def\sq{\hbox{\rlap{$\sqcap$}$\sqcup$}}

\def\bbbone{{\mathchoice {\rm 1\mskip-4mu l} {\rm 1\mskip-4mu l}
{\rm 1\mskip-4.5mu l} {\rm 1\mskip-5mu l}}}

\def\ind{\bbbone}

\usepackage{amsmath} 
\usepackage{amsfonts}
\usepackage{latexsym}
\usepackage{bm}

\newcommand{\ben}{\begin{enumerate}}
\newcommand{\een}{\end{enumerate}}
\newcommand{\bit}{\begin{itemize}}
\newcommand{\eit}{\end{itemize}}

\newtheorem{theorem}{Theorem}[section]
\newtheorem{proposition}[theorem]{Proposition}

\newcommand{\ba}{\begin{array}{rcl}}
\newcommand{\ea}{\end{array}}
\newcommand{\bt}{\begin{theorem}}
\newcommand{\et}{\end{theorem}}
\newcommand{\bd}{\begin{description}}
\newcommand{\ed}{\end{description}}

\def\Expect{{\sf E}}

\def\slabel#1{\label{s:#1}}

\def\elabel#1{\label{e:#1}}
\def\eq#1/{(\ref{e:#1})}

\def\Section#1{Section~\ref{s:#1}}

\def\beq{\begin{equation}}
\def\eeq{\end{equation}}
\def\beqa{\begin{eqnarray}}
\def\eeqa{\end{eqnarray}}

\def\qed{\ifmmode\sq\else{\unskip\nobreak\hfil
\penalty50\hskip1em\null\nobreak\hfil\sq
\parfillskip=0pt\finalhyphendemerits=0\endgraf}\fi}

\def\sqr#1#2{{\vcenter{\hrule height.#2pt
      \hbox{\vrule width.#2pt height#1pt \kern#1pt
         \vrule width.#2pt}
       \hrule height.#2pt}}}

\newcommand{\bc}{\begin{corollary}}
\newcommand{\ec}{\end{corollary}}

\newcommand{\bp}{\begin{proposition}}
\newcommand{\ep}{\end{proposition}}

\def\eye(#1){{\bf (#1)}\quad}

\def\taboo#1{{{}_{#1}}}

\def\0P{\taboo{0}P}
\def\0Pn{\taboo{0}P^n}

\def\divline{\begin{center} \line(1,0){270} \end{center}}

\usepackage{amsmath}
\usepackage{bbm}
\DeclareSymbolFont{largesymbols}{OMX}{yhex}{m}{n}
\DeclareMathAccent{\widewidehat}{\mathord}{largesymbols}{"62}

\usepackage{algorithm}
\usepackage{algorithmic}
\usepackage{color}
\usepackage{graphicx}
\usepackage[position=t,singlelinecheck=off]{subfig}
\usepackage{caption}
\usepackage[hang,flushmargin]{footmisc}

\date{July 9, 2014}

\title{Particle Filtering and Smoothing Using Windowed Rejection Sampling}

\def\jem{Postal Address:
      Department of Applied Mathematics,
      University of Colorado, Box 526
      Boulder CO 80309-0526, USA; email: corcoran@colorado.edu, 
      phone: 303-492-0685}

\author {J. N. Corcoran and D. Jennings\\
University of Colorado \thanks{\jem} }

\begin{document}

%\pagebreak
%\tableofcontents
%\pagebreak

\maketitle
\vspace{-.8cm}

\begin{abstract} \small \noindent
``Particle methods" are sequential Monte Carlo algorithms, typically involving importance sampling, that are used to estimate and sample from joint and marginal densities from a collection of a, presumably increasing, number of random variables. In particular, a particle filter aims to estimate the current state $X_{n}$ of a stochastic system that is not directly observable by estimating a posterior distribution $\pi(x_{n}|y_{1},y_{2}, \ldots, y_{n})$ where the $\{Y_{n}\}$ are observations related to the $\{X_{n}\}$ through some measurement model $\pi(y_{n}|x_{n})$. A particle smoother aims to estimate a marginal distribution $\pi(x_{i}|y_{1},y_{2}, \ldots, y_{n})$ for $1 \leq i < n$. Particle methods are used extensively for hidden Markov models where $\{X_{n}\}$ is a Markov chain as well as for more general state space models.

\noindent Existing particle filtering algorithms are extremely fast and easy to implement. Although they suffer from issues of degeneracy and ``sample impoverishment", steps can be taken to minimize these problems and overall they are excellent tools for inference. However, if one wishes to sample from a posterior distribution of interest, a particle filter is only able to produce dependent draws. Particle smoothing algorithms are complicated and far less robust, often requiring cumbersome post-processing, ``forward-backward" recursions, and multiple passes through subroutines. In this paper we introduce an alternative algorithm for both filtering and smoothing that is based on rejection sampling ``in windows" . We compare both speed and accuracy of the traditional particle filter and this ``windowed rejection sampler" (WRS)  for several examples and show that good estimates for smoothing distributions are obtained at no extra cost.

\bigskip

\end{abstract}


\setcounter{page}{0}

%\tableofcontents

%%%%%%%%%%%%%%%%%%%%%%%%%%%%%%%%%%%%%%%%%%%%%%%%%%%%%%%%%%%%%%%%%%%%%%%%%
%%%%%%%%%%%%%%%%%%%%%%%      INTRODUCTION       %%%%%%%%%%%%%%%%%%%%%%%%%
%%%%%%%%%%%%%%%%%%%%%%%%%%%%%%%%%%%%%%%%%%%%%%%%%%%%%%%%%%%%%%%%%%%%%%%%%

\section{Introduction}
\slabel{int} 
Particle filters and smoothers are sequential Monte Carlo methods, typically importance sampling methods, that are often employed to sample from and provide estimates of the distribution of a set or subset of latent variables in a hidden Markov model given observations. They are constructed specifically to provide updated sampling and estimation when additional observations become available without reprocessing all observations.

\begin{figure}[!ht]
  \captionsetup[subfigure]{labelformat=simple}
  \centering
  \caption{Depiction of One and Two Layer Hidden Markov Models}
  \label{fig:hmm}
  \subfloat[]{\includegraphics[width=.4\textwidth]{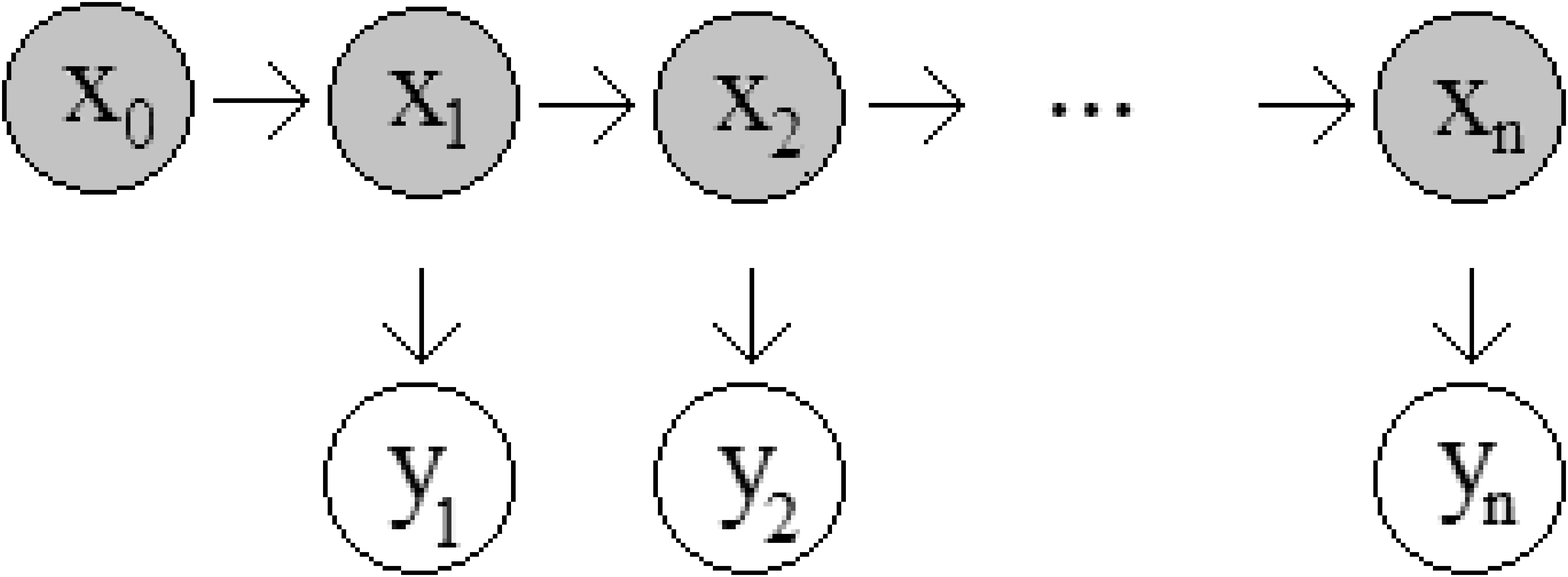}}\qquad
  \subfloat[]{\includegraphics[width=.4\textwidth]{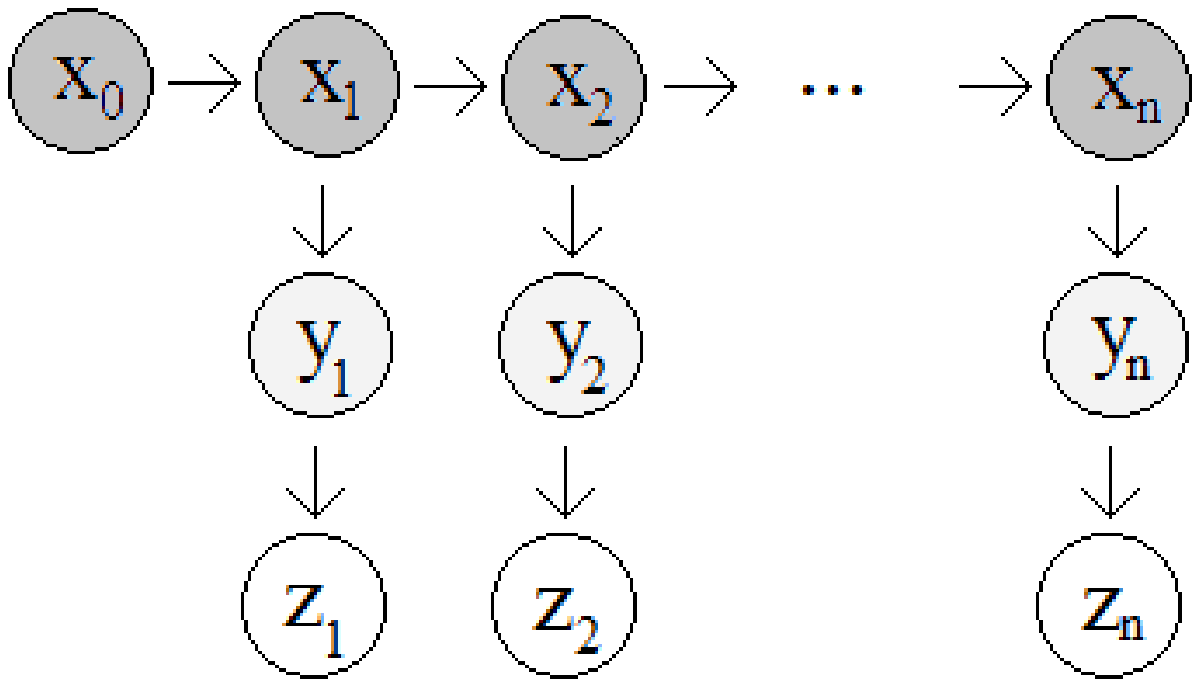}}\\
\vspace{-0.1in}
(a) \qquad \qquad \qquad \qquad \qquad \qquad \qquad \qquad \qquad  (b)
\end{figure}

Consider the hidden Markov model with underlying and unobserved states $\{X_{n}\}_{n=0}^{\infty}$, transition density $\pi(x_{n}|x_{n-1})$, and an initial distribution with density $\pi(x_{0})$. (In this paper we will assume a continuous state space, though the sampling techniques described will apply in the discrete case as well.) Suppose that $\{Y_{n}\}_{n=1}^{\infty}$ represents a sequence of observable variables that are conditionally independent when the unobserved states are given and where each $Y_{n}$ is related to the unobserved process through $X_{n}$ and a ``measurement model" density $\pi(y_{n}|x_{n})$. Such a model is depicted in Figure \ref{fig:hmm}(a).

Our goal is to sample from the density $\pi(x_{0:n}|y_{1:n})$, where $u_{i:j}$ denotes the vector $(u_{i}, u_{i+1}, \ldots, u_{j})$ for $j \geq i$, sequentially in the sense that samples from $\pi(x_{0:n-1}|y_{1:n-1})$ will be used along with a new observation $y_{n}$ to produce the desired points. The sampled points can then be used to approximate, for example, expectations of the form $\Expect[f(X_{0:n})|y_{1:n}]$ and  marginal distributions $\pi(x_{i}|y_{1:n})$ for $i=0,1,\ldots,n$. The estimation of $\pi(x_{n}|y_{1:n})$ is known as the {\emph{filtering}} problem and that of $\pi(x_{i}|y_{1:n})$ for $0 \leq i < n$ is known as the {\emph{smoothing}} problem. In \Section{examples}, we will see that our proposed algorithm, based on rejection sampling ``in windows'', vastly outperforms traditional particle methods for the smoothing problem and has some advantages for filtering as well.

%In examples to follow, the chain $\{X_{n}\}$ may start at time $0$ or time $1$, and, in the case where we have $X_{0}$, our $\{Y_{n}\}$ observations may start at either time $0$ or $1$. Furthermore, we will also consider a simple two-layer hidden Markov model, as depicted in Figure \ref{fig:hmm}(b), where only $\{Z_{n}\}$ is observed for $n \geq 0$ or $n \geq 1$.

The Markov and conditional independence assumptions allow us to write our ``target density" for the model in Figure \ref{fig:hmm}(a) in the recursive form

\begin{subequations}
\beqa
%\elabel{target}
\pi(x_{0:n}|y_{1:n}) &\propto&  \pi(x_{0}) \prod_{i=1}^{n} \pi(y_{i}|x_{i}) \pi(x_{i}|x_{i-1}) \elabel{targeta} \\
\nonumber \\
&\propto & \pi(y_{n}|x_{n}) \cdot \pi(x_{n}|x_{n-1}) \cdot \pi(x_{0:n-1}|y_{1:n-1}) \elabel{targetb} 
\eeqa
\end{subequations}
so that draws from $\pi(x_{0:n}|y_{1:n})$ can be related to draws from $\pi(x_{0:n-1}|y_{1:n-1})$.

In \Section{exist}, we review the sequential importance sampling and resampling methods, which make up the most commonly used particle filtering algorithm, that is typically used to estimate and sample from (\ref{e:targeta}). We will then offer our new rejection sampling alternative in \Section{wrs}. In \Section{examples} we give simulation results, including direct comparisons to traditional particle filtering and an application to a second hidden layer Markov model.

\section{Importance Sampling Based Sequential Methods}
\slabel{exist} 

We now briefly describe existing importance sampling and resampling approaches to the problem of estimating and/or sampling sequentially from a target density
\beq
\elabel{target2}
\pi(x_{1:n}) = h(x_{1:n})/Z_{n},
\eeq
where $Z_{n} = \int h(x_{1:n}) \, dx_{1:n}$. A much more complete review is given by Doucet and Johansen \cite{doucet2008}. Without loss of generality, $x_{1:n}$ can be replaced by $x_{0:n}$ and the target density can be a conditional density.

To establish notation, note that, if $X_{1:n}^{(1)}, X_{1:n}^{(2)}, \ldots, X_{1:n}^{(N)}$ represent $N$ points (``particles'') sampled from $\pi(x_{1:n})$, sequentially or otherwise, a simple unbiased estimator of the target density is given by the empirical density that puts weight $1/N$ on each of the points. We may write this succinctly as
\beq
\elabel{est1}
\widehat{\pi}(x_{1:n}) = \frac{1}{N} \sum_{i=1}^{N} \ind [X_{1:n}^{(i)} =x_{1:n}]  
\eeq
where $\ind [X_{1:n}^{(i)} =x_{1:n}]$ is the indicator function that takes on
the value $1$ when $x_{1:n} = X_{1:n}^{(i)}$, and zero otherwise.

We can then estimate, for example, $\Expect[f(X_{1:n})]$, for a given function $f$ by
\beq
\elabel{expfunc}
\widehat{\Expect}[f(X_{1:n})]= \frac{1}{N} \sum_{i=1}^{N} f(X_{1:n}^{(i)}).
\eeq

It is easy to verify that (\ref{e:est1}) and (\ref{e:expfunc}) are unbiased for $\pi(x_{1:n})$ and $\Expect[f(X_{1:n})]$, respectively.

\subsection{A Review of Non-Sequential Importance Sampling}
\slabel{importance}
When one can not sample directly from $\pi(x_{1:n}) \propto h(x_{1:n})$, importance sampling can be used to instead sample points from a more tractable density, and these points can be used to estimate the target density. Suppose that $q(x_{1:n})$ is another density, presumably tractable, with the same support as $\pi(x_{1:n})$.

We can write 
\beq
\elabel{pwithq}
\pi(x_{1:n}) = \frac{h(x_{1:n})}{Z_{n}} = \frac{w(x_{1:n})q(x_{1:n})}{Z_{n}}
\eeq
where
\beq
\elabel{weights}
w(x_{1:n}) := \frac{h(x_{1:n})}{q(x_{1:n})}.
\eeq
Then we proceed to estimate $\pi(x_{1:n})$ and $Z_{n}$ as follows.
\begin{itemize}
\item Sample $X_{1:n}^{(1)}, X_{1:n}^{(2)}, \ldots, X_{1:n}^{(N)} \stackrel{iid}{\sim } q(x_{1:n})$.\footnote{``iid'' denotes  ``independent and identically distributed'' values.}
\item Estimate $q(x_{1:n})$ with
\beq
\elabel{estq}
\widehat{q(x_{1:n})} = \frac{1}{N} \sum_{i=1}^{N} \ind [X_{1:n}^{(i)}=x_{1:n}]  
\eeq 
\item Estimate $Z_{n} = \int w(x_{1:n}) q(x_{1:n}) \, dx_{1:n} = \Expect[w(X_{1:n})]$, when $X_{1:n} \sim q(x_{1:n})$, with
\beq
\elabel{zest}
\widehat{Z_{n}} = \frac{1}{N} \sum_{i=1}^{N} w(X_{1:n}^{(i)}).
\eeq
\item Estimate $\pi(x_{1:n})$ with
\beq
\elabel{est2}
\begin{array}{lcl}
\widehat{\pi(x_{1:n})} &=&\frac{w(x_{1:n}) \widehat{q(x_{1:n})}}{\widehat{Z_{n}}} =\frac{\frac{1}{N} \sum_{i=1}^{N} w(X_{1:n}^{(i)}) \ind [X_{1:n}^{(i)}=x_{1:n}]}{\widehat{Z_{n}}}\\
\\
& =& \sum_{i=1}^{N} W_{n}^{(i)} \, \ind [X_{1:n}^{(i)}=x_{1:n}]  
\end{array}
\eeq 
where
$$
W_{n}^{(i)} = \frac{w(X_{1:n}^{(i)})}{\sum_{j=1}^{N} w(X_{1:n}^{(j)})}.
$$
\end{itemize}

It is routine to show that (\ref{e:est2}) is an unbiased estimator of $\pi(x_{1:n})$ and that the optimal importance sampling density $q(x_{1:n})$, in terms of minimizing variance,  is $q(x_{1:n}) \equiv \pi(x_{1:n})$.

\subsection{Sequential Importance Sampling (SIS)}
\slabel{sis}
Sequential importance sampling (SIS) is importance sampling for $\pi(x_{1:n})$ in such a way where draws from $\pi(x_{1:n-1})$ are ``extended" to $n$-dimensional points that are then reweighted to produce draws from $\pi(x_{1:n})$.  To this end, the importance sampling density is chosen to have the form
$$
q(x_{1:n}) = q(x_{1}) \prod_{i=2}^{n} q(x_{i}|x_{i-1})
$$
so that it may be sampled from recursively.

The associated importance sample weights may then also be computed recursively since, for $n \geq 2$,
\beqa
\elabel{recursiveweights}
\nonumber w(x_{1:n}) &=& \frac{h(x_{1:n})}{q(x_{1:n})} = \frac{h(x_{1:n})}{q(x_{1:n-1})  q(x_{n}|x_{n-1})}\\
\nonumber \\
\nonumber &=& \frac{h(x_{1:n-1})}{q(x_{1:n-1})} \cdot \frac{h(x_{1:n})}{h(x_{1:n-1}) q(x_{n}|x_{n-1})}\\
\nonumber \\
&=: &w(x_{1:n-1}) \cdot \alpha(x_{1:n})
\eeqa
where $\alpha(x_{1:n})$ is an \emph{incremental weight function} that is defined as
$$
\alpha(x_{1:n}) := \frac{h(x_{1:n})}{h(x_{1:n-1}) q(x_{n}|x_{n-1})}. 
$$

\divline
\vspace{0.2in}
{\bf{\large SIS Algorithm}}

At the first time step ($n=1$),
\begin{itemize}
\item Sample $X_{1}^{(1)}, X_{1}^{(2)}, \ldots, X_{1}^{(N)} \stackrel{iid}{\sim} q(x_{1})$.
\item Compute weights $w(X_{1}^{(i)})$ for $i=1,2,\ldots, N$ using (\ref{e:weights}).
\end{itemize}

At times $n \geq 2$,
\begin{itemize}
\item Sample $X_{n}^{(1)}, X_{n}^{(2)}, \ldots, X_{n}^{(N)}$ independently with $X_{n}^{(i)} \sim  q(x_{n}|X_{n-1}^{(i)})$.
\item Compute weights $w(X_{1:n}^{(i)}) = w(X_{1:n-1}^{(i)}) \cdot \alpha(X_{1:n}^{(i)})$ for $i=1,2,\ldots, N$.
\end{itemize}

\divline
At any time $n$, one can estimate $\pi(x_{1:n})$ and $Z_{n}$ using (\ref{e:est2}) and (\ref{e:zest}). One can also obtain approximate dependent draws from $\pi(x_{1:n})$ by sampling from (\ref{e:est2}). That is, by sampling from the set of values $\{X_{1:n}^{(1)}, X_{1:n}^{(2)}, \ldots, X_{1:n}^{(N)}\}$ using respective weights $\{W_{n}^{(1)}, W_{n}^{(2)}, \ldots, W_{n}^{(N)}\}$.

In practice, iteration of the SIS algorithm leads to a ``degeneracy of weights'' problem (see, for example, \cite{anddoucethol}, \cite{doucet2001}, \cite{doucetgod},  and \cite{haug2012}) where the weights of all but one particle will approach zero, causing the method to break down and give meaningless results.

%\divline
%\vspace{0.1in}
One way to avoid the issue of degenerate weights is to implement a resampling scheme at each time step.

\divline

\vspace{0.1in}
{\bf{\large SIS Algorithm With Resampling (SIR)}}

At the first time step ($n=1$),
\begin{itemize}
\item Sample $X_{1}^{(1)}, X_{1}^{(2)}, \ldots, X_{1}^{(N)} 
\stackrel{iid}{\sim} q(x_{1})$.
\item Compute weights $w(X_{1}^{(i)})$ for $i=1,2,\ldots, N$.
\item Compute the normalized weights
$$
W_{1}^{(i)} = \frac{w(X_{1}^{(i)})}{\sum_{j=1}^{N} w(X_{1}^{(j)})}
$$
for $i=1,2,\ldots, N$.
\item Sample $N$ points, $\widetilde{X}_{1}^{(1)}, \widetilde{X}_{1}^{(2)}, \ldots, \widetilde{X}_{1}^{(N)}$, with replacement,  from the set $\{X_{1}^{(1)}, X_{1}^{(2)}, \ldots, X_{1}^{(N)}\}$ with respective probabilities $\{W_{1}^{(1)}, W_{1}^{(2)}, \ldots, W_{1}^{(N)}\}$. That is, sample $\widetilde{X}_{1}^{(1)}, \widetilde{X}_{1}^{(2)}, \ldots, \widetilde{X}_{1}^{(N)}$ from 
$$
\widehat{\pi(x_{1})} = \sum_{i=1}^{N} W_{1}^{(i)} \, \ind [X_{1}^{(i)}=x_{1}].
$$

Note that  $\widetilde{X}_{1}^{(1)}, \widetilde{X}_{1}^{(2)}, \ldots, \widetilde{X}_{1}^{(N)}$ are now equally weighted particles, each with weight $1/N$. 

Assign weights $w(\widetilde{X}_{1}^{(i)})=1/N$ for $i=1,2,\ldots, N$.

\end{itemize}

At times $n \geq 2$,
\begin{itemize}
\item Sample $X_{n}^{(1)}, X_{n}^{(2)}, \ldots, X_{n}^{(N)}$ independently 
with $X_{n}^{(i)} \sim  q(x_{n}|\widetilde{X}_{n-1}^{(i)})$.

\item Extend each particle $\widetilde{X}_{1:n-1}^{(i)}$ to particles $(\widetilde{X}_{1:n-1}^{(i)},X_{n}^{(i)})$.

\item Compute associated weights $w(\widetilde{X}_{1:n-1}^{(i)},X_{n}^{(i)}) := \alpha(\widetilde{X}_{1:n-1}^{(i)},X_{n}^{(i)})$ for $i=1,2,\ldots, N$. 

(This is consistent with SIS since the previous weights in (\ref{e:recursiveweights}) have been replaced by the constant $1/N$ and the current weights have yet to be normalized.) 

\item Compute the normalized weights
$$
W_{n}^{(i)} = \frac{w(\widetilde{X}_{1:n-1}^{(i)},X_{n}^{(i)})}{\sum_{j=1}^{N} w(\widetilde{X}_{1:n-1}^{(j)},X_{n}^{(j)})}
$$
for $i=1,2,\ldots, N$.

\item Sample $N$ $n$-dimensional points, $\widetilde{X}_{1:n}^{(1)}, \widetilde{X}_{1:n}^{(2)},
 \ldots, \widetilde{X}_{1:n}^{(N)}$, with replacement,  from the set 
$\{(\widetilde{X}_{1:n-1}^{(1)},X_{n}^{(i)}), (\widetilde{X}_{1:n-1}^{(2)},X_{n}^{(2)}), \ldots, (\widetilde{X}_{1:n-1}^{(N)},X_{n}^{(N)})\}$ with respective 
probabilities $\{W_{n}^{(1)}, W_{n}^{(2)}, \ldots, W_{n}^{(N)}\}$. That is, sample $\widetilde{X}_{1:n}^{(1)}, \widetilde{X}_{1:n}^{(2)}, \ldots, \widetilde{X}_{1:n}^{(N)}$ from 
\beq
\elabel{sirestimator1}
\widehat{\pi(x_{1:n})} = \sum_{i=1}^{N} W_{n}^{(i)} \, \ind [(\widetilde{X}_{1:n-1}^{(i)},X_{n}^{(i)})=x_{1:n}].
\eeq

Note that  $\widetilde{X}_{1:n}^{(1)}, \widetilde{X}_{1:n}^{(2)}, \ldots, 
\widetilde{X}_{1:n}^{(N)}$ are now equally weighted particles, each with 
weight $1/N$.

Assign weights $w(\widetilde{X}_{1:n}^{(i)})=1/N$ for $i=1,2,\ldots, N$.
\end{itemize}

\divline
As with SIS, we may, at any time $n$, estimate $\pi(x_{1:n})$ using (\ref{e:sirestimator1}) or by using 
%$\widetilde{X}_{1:n}^{(1)}, \widetilde{X}_{1:n}^{(2)}, \ldots, \widetilde{X}_{1%:n}^{(N)}$ with (\ref{e:est1}). 
\beq
\elabel{sirestimator2}
\widehat{\pi(x_{1:n})} = \frac{1}{N}\sum_{i=1}^{N} \ind [\widetilde{X}_{1:n}^{(i)}=x_{1:n}].
\eeq
We may obtain approximate dependent draws from 
$\pi(x_{1:n})$ by sampling from (\ref{e:sirestimator2}). That is, by sampling uniformly, with replacement, from the set of values $\{\widetilde{X}_{1:n}^{(1)}, \widetilde{X}_{1:n}^{(2)}, \ldots, \widetilde{X}_{1:n}^{(N)}\}$.

An obvious issue with the SIR algorithm, known as the {\emph{sample impoverishment problem}}, is a decreasing  diversity of particles as those with higher weights will likely be drawn multiple times. Additionally, the resampling step may make the algorithm prohibitively slow and can lead to increased variance of estimates. One way to address the speed and variance is to only resample in the case of a small (user chosen threshold for)  {\emph{effective sample size}}
\beq
\elabel{eff}
N_{eff} := \left[ \sum_{i=1}^{N} (W_{n}^{(i)})^{2} \right]^{-1}.
\eeq
A small effective sample size indicates higher variability of the weights which, in turn, indicates impending degeneracy. In this case, resampling would be especially prudent.

It is important to note that, when resampling, one should be originally sampling a very large number of points in order to minimize the effects of resampling repeated values from a discrete distribution.
 Additional details about SIS and SIR procedures  can be found in Doucet \emph{at al.} \cite{doucet2001,doucetgod} and in Andrieu and Doucet \cite{anddouc}.  
%There are suggested techniques for dealing with this issue, as well as degeneracy, in  \cite{anddouc, doucet2001,doucetgod}.  

\section{The Windowed Rejection Sampler}
\slabel{wrs}

In order to draw from a possibly unnormalized density $\pi(x) \propto h(x)$ using rejection sampling, one must find a density $q(x)$ and a constant $M$ such that
$$
h(x) \leq M q(x).
$$
Then, one proceeds as in Algorithm \ref{alg1}.
\begin{algorithm}
\caption{{\bf{Rejection Sampling from \boldmath{$\pi(x)$}:}}}
\label{alg1}
\begin{algorithmic}
\STATE 1. Generate $X \sim q$.
\STATE 2. Generate $U \sim \mbox{Uniform}(0,1)$.
\STATE 3. If $U > \frac{h(X)}{Mq(X)}$, return to 1. Otherwise, if $U \leq \frac{h(X)}{Mq(X)}$, $X$ is a draw from $\pi$. 
\end{algorithmic}
\end{algorithm}

Since $\pi(x)$ may be a high-dimensional and conditional density, we could, in theory, use rejection sampling to draw exactly from
$$
\begin{array}{lcl}
\pi(x_{0:n}|y_{1:n}) \propto h(x_{0:n}|y_{1:n}) &:=& \pi(x_{0}) \prod_{i=1}^{n} \pi(y_{i}|x_{i}) \pi(x_{i}|x_{i-1}) \\
\\
&=&\underbrace{\left[ \prod_{i=1}^{n} \pi(y_{i}|x_{i}^{*})  \right]}_{M} \cdot \underbrace{\pi(x_{0}) \prod_{i=1}^{n}  \pi(x_{i}|x_{i-1})}_{q(x_{0:n})}
\end{array}
$$
where
$$
x_{i}^{*} = \arg\max_{x_{i}} \pi(y_{i}|x_{i}).
$$
For most applications of particle filtering in the literature, $x_{i}^{*}$, or at least an upper bound on $\pi(y_{i}|x_{i})$, with respect to $x_{i}$, is easily attainable. However, for even small values of $n$, the acceptance probability, $\prod_{i=1}^{n} \pi(y_{i}|x_{i})/\pi(y_{i}|x_{i}^{*})$, can be prohibitively small and of course we are losing the benefit of sequential sampling. Thus, we propose rejection sampling ``in windows". 

\subsection*{Rejection Sampling in Windows}

``Windowed Rejection Sampling" (WRS) is based on the idea that, depending on the covariance structure of the chain, at some point future observed $y$'s will eventually cease to significantly affect earlier $x$'s. For example, if $\pi(x_{0}|y_{1},y_{2},y_{3},y_{4}) \approx \pi(x_{0}|y_{1},y_{2},y_{3})$, we can write
$$
\begin{array}{lcl}
\pi(x_{0},x_{1},x_{2},x_{3},x_{4}|y_{1},y_{2},y_{3},y_{4}) &=& \pi(x_{1},x_{2},x_{3},x_{4}|x_{0},y_{1},y_{2},y_{3},y_{4}) \cdot \pi(x_{0}|y_{1},y_{2},y_{3},y_{4})\\
\\
&\approx& \pi(x_{1},x_{2},x_{3},x_{4}|x_{0},y_{1},y_{2},y_{3},y_{4}) \cdot \pi(x_{0}|y_{1},y_{2},y_{3}),
\end{array}
$$
and so we can sample approximately from $\pi(x_{0:4}|y_{1:4})$ by first sampling $X_{0}$ from $\pi(x_{0}|y_{1:3})$ and then sampling $X_{1:4}$ from 
$\pi(x_{1},x_{2},x_{3},x_{4}|x_{0},y_{1},y_{2},y_{3},y_{4})$. Sampling from $\pi(x_{0}|y_{1:3})$ can be achieved by sampling $X_{0:3}$ from $\pi(x_{0:3}|y_{1:3})$ and considering only the $x_{0}$ values. In this example, we say that we are using rejection sampling in a ``window of length 4".

More formally, the WRS algorithm is described in Algorithm \ref{wrsalg} for a given window length $w$. We discuss the choice of this tuning parameter in Sections 4 and 5. 

\begin{minipage}{\textwidth}
\renewcommand\footnoterule{} 
\begin{algorithm}[H]
\caption{{\bf{Windowed Rejection Sampling from \boldmath{$\pi(x_{0:n}|y_{1:n})$}:}}}
\label{wrsalg}
\begin{algorithmic}
\STATE Find $x_{i}^{*} = \arg\max_{x_{i}} \pi(y_{i}|x_{i})$, and set a window length parameter $w$. \footnotetext{\noindent Note that if it is not obtainable, $\pi(y_{i}|x_{i}^{*})$ in expressions for $M$ can be replaced by any upper bound.}%
\vspace{0.1in}
\STATE 1. Generate $X_{0:w-1} \sim \pi(x_{0:w-1}|y_{1:w-1})$ using rejection sampling with 
$$q(x_{0:w-1}) = \pi(x_{0}) \prod_{i=1}^{w-1} \pi(x_{i}|x_{i-1})$$
and $M = \prod_{i=1}^{w} \pi(y_{i}|x_{i-1}^{*})$. 
Set $m=1$.
\vspace{0.1in}
\STATE 2. Generate $X_{m:m+w-1} \sim \pi(x_{m:m+w-1}|x_{m-1},y_{1:m+w-1})=\pi(x_{m:m+w-1}|x_{m-1},y_{m:m+w-1})$ using rejection sampling with
$$
q(x_{m:m+w-1}|x_{m-1}) = \prod_{i=m}^{m+w-1} \pi(x_{i}|x_{i-1})
$$
and $M=\prod_{i=m}^{m+w-1} \pi(y_{i}|x_{i}^{*})$.

Note that
$$
\begin{array}{lcl}
\pi(x_{m:m+w-1}|x_{m-1},y_{m:m+w-1}) &\propto& h(x_{m:m+w-1}|x_{m-1},y_{m:m+w-1}) \\
\\
&:= &\prod_{i=m}^{m+w-1} \pi(y_{i}|x_{i}) \pi(x_{i}|x_{i-1}) \leq M \cdot q(x_{m,n+w-1}|x_{m-1}).
\end{array}
$$

\vspace{0.1in}
Set $m=m+1$ and return to the beginning of Step 2. Continue until $m+w-1=n$.
\end{algorithmic}
\end{algorithm}
\end{minipage}
\vspace{0.3in}

\section{Examples}
\slabel{examples}

\subsection{One Dimensional Normals with One Hidden Layer}
\slabel{normalex}
We begin, for the purpose of illustration, with a very explicit description of the algorithm with a window length of $w=3$, for the simple model
$$
X_{0} \sim N(\mu_{0},\sigma_{0}^{2}), \qquad
X_{n+1} = aX_{n} + \sigma_{X} \varepsilon_{n+1}, \qquad
Y_{n}= b X_{n} + \sigma_{Y} \nu_{n}
$$
where $\{\varepsilon_{n}\} \stackrel{iid}{\sim} N(0,1)$ and $\{\nu_{n}\} \stackrel{iid}{\sim} N(0,1)$ are independent sequences. Assume that only the $Y_{n}$ are observed.

Let $N(x;\mu,\sigma^{2})$ denote the $N(\mu,\sigma^{2})$ density.

We begin by using rejection sampling to produce $N$ iid draws from $\pi(x_{0:2}|y_{1:2})$
$$
\pi(x_{0:2}|y_{1:2})  \propto [\pi(y_{1}|x_{1}) \pi(y_{2}|x_{2}) ] \cdot [\pi(x_{2}|x_{1}) \pi(x_{1}|x_{0}) \pi(x_{0})]
$$
using $q(x_{0:2}) = \pi(x_{2}|x_{1}) \pi(x_{1}|x_{0}) \pi(x_{0}) = N(x_{2};ax_{1},\sigma_{X}^{2}) \cdot N(x_{1}; ax_{0},\sigma_{X}^{2}) \cdot N(x_{0};\mu_{0},\sigma_{0}^{2})$.

It is easy to see that 
$$
x_{i}^{*} = \arg\max_{x_{i}} \pi(y_{i}|x_{i}) = \arg\max_{x_{i}} N(y_{i};bx_{i},\sigma_{Y}^{2})=y_{i}/b \qquad \mbox{for} \,\,\, i=1,2,\ldots.
$$
We repeatedly draw independent realizations $x_{0:2}$ of $X_{0:2} \sim q(x_{0:2})$ and $u$ of $U \sim \mbox{Uniform}(0,1)$ until the first time that 
$$
u \leq \frac{\pi(y_{1}|x_{1}) \pi(y_{2}|x_{2}) }{\pi(y_{1}|x_{1}^{*}) \pi(y_{2}|x_{2}^{*}) }.
$$
Repeating this procedure $N$ times, 
we collect only the values of $x_{0}$ and record them as $X_{0}^{(1)}, X_{0}^{(2)}, \ldots, X_{0}^{(N)}$.

Now, moving our window of length $w=3$ to the right by $1$, 
we use rejection sampling to produce $N$ iid draws from $\pi(x_{1:3}|x_{0},y_{1:3})$
$$
\pi(x_{1:3}|x_{0},y_{1:3})  \propto [\pi(y_{1}|x_{1}) \pi(y_{2}|x_{2}) \pi(y_{3}|x_{3})] \cdot [ \pi(x_{3}|x_{2}) \pi(x_{2}|x_{1}) \pi(x_{1}|x_{0})]
$$
using 
$$q(x_{1:3}|x_{0}) = \pi(x_{3}|x_{2}) \pi(x_{2}|x_{1}) \pi(x_{1}|x_{0}) = N(x_{3};ax_{2},\sigma_{X}^{2}) \cdot N(x_{2}; ax_{1},\sigma_{X}^{2}) \cdot N(x_{1}; ax_{0},\sigma_{X}^{2}). $$
That is, for each $i=1,2, \ldots, N$, we repeatedly draw independent realizations $x_{1:3}$ of $X_{1:3} \sim q(x_{1:3}|X_{0}^{(i)})$ and $u$ of $U \sim \mbox{Uniform}(0,1)$ until the first time that 
$$
u \leq \frac{\pi(y_{1}|x_{1}) \pi(y_{2}|x_{2}) \pi(y_{3}|x_{3})}{\pi(y_{1}|x_{1}^{*}) \pi(y_{2}|x_{2}^{*}) \pi(y_{3}|x_{3}^{*})},
$$
collecting the resulting value of $x_{1}$ and recording it as $X_{1}^{(i)}$.

Move the window of length $w=3$ to the right by $1$ to produce $N$ draws from $\pi(x_{2:4}|x_{1},y_{1:4}) = \pi(x_{2:4}|x_{1},y_{2:4})$, and retain the resulting values for $X_{2}^{(i)}$ for $i=1,2,\ldots,N$.

\begin{figure}[!ht]
  \captionsetup[subfigure]{labelformat=simple}
  \centering
  \caption{Section 4.1 Example: Means for marginal components in $100,000$ draws from $\pi(x_{0:10}|y_{1:10})$ using perfect 11-dimensional rejection sampling, the SIR algorithm, and the WRS algorithm with various window lengths.}
  \label{fig:example1}
\vspace{-0.3in}
\includegraphics[width=4in,height=2.9in]{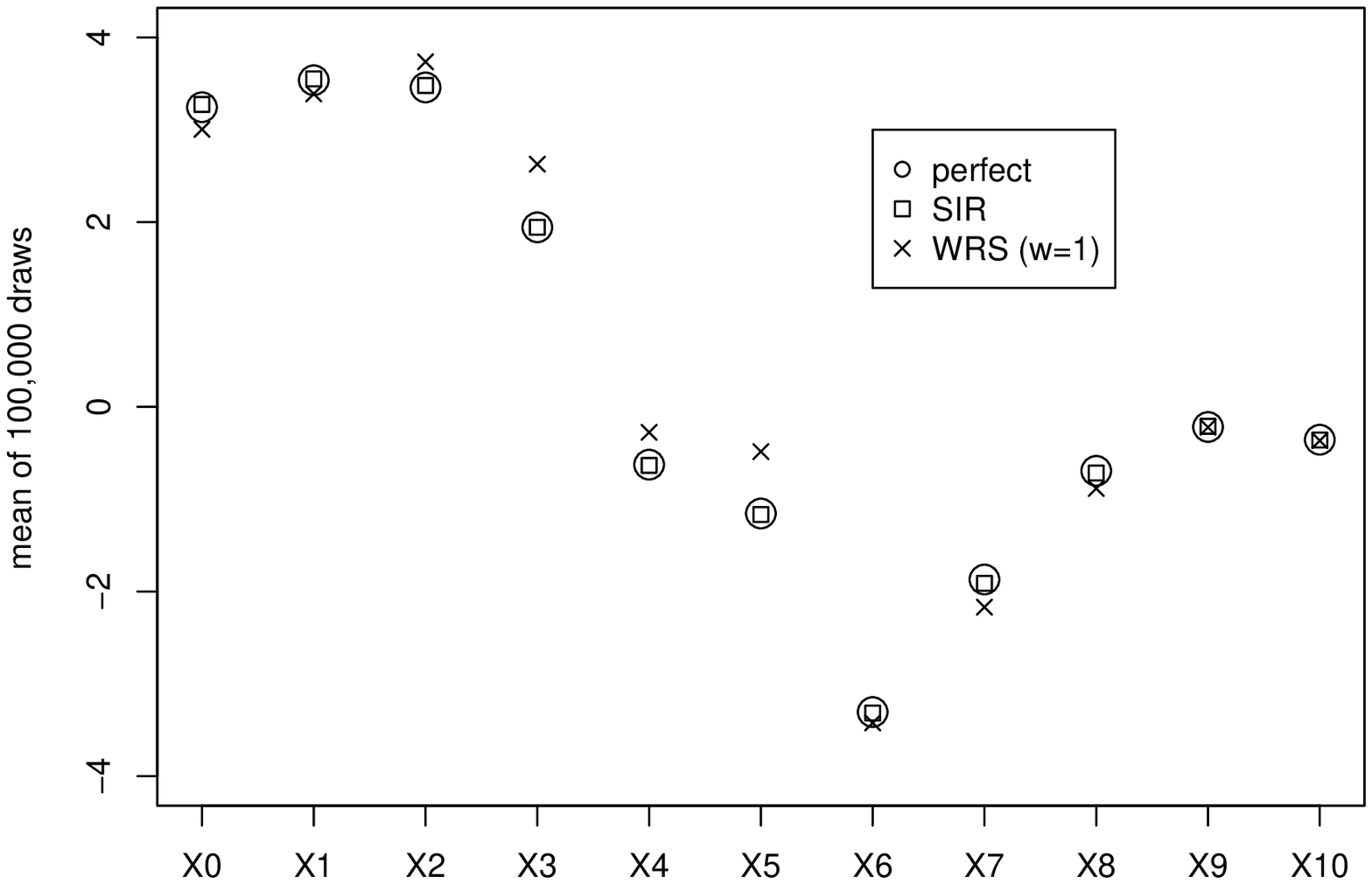}\\
\vspace{-0.1in}
\includegraphics[width=4in,height=2.9in]{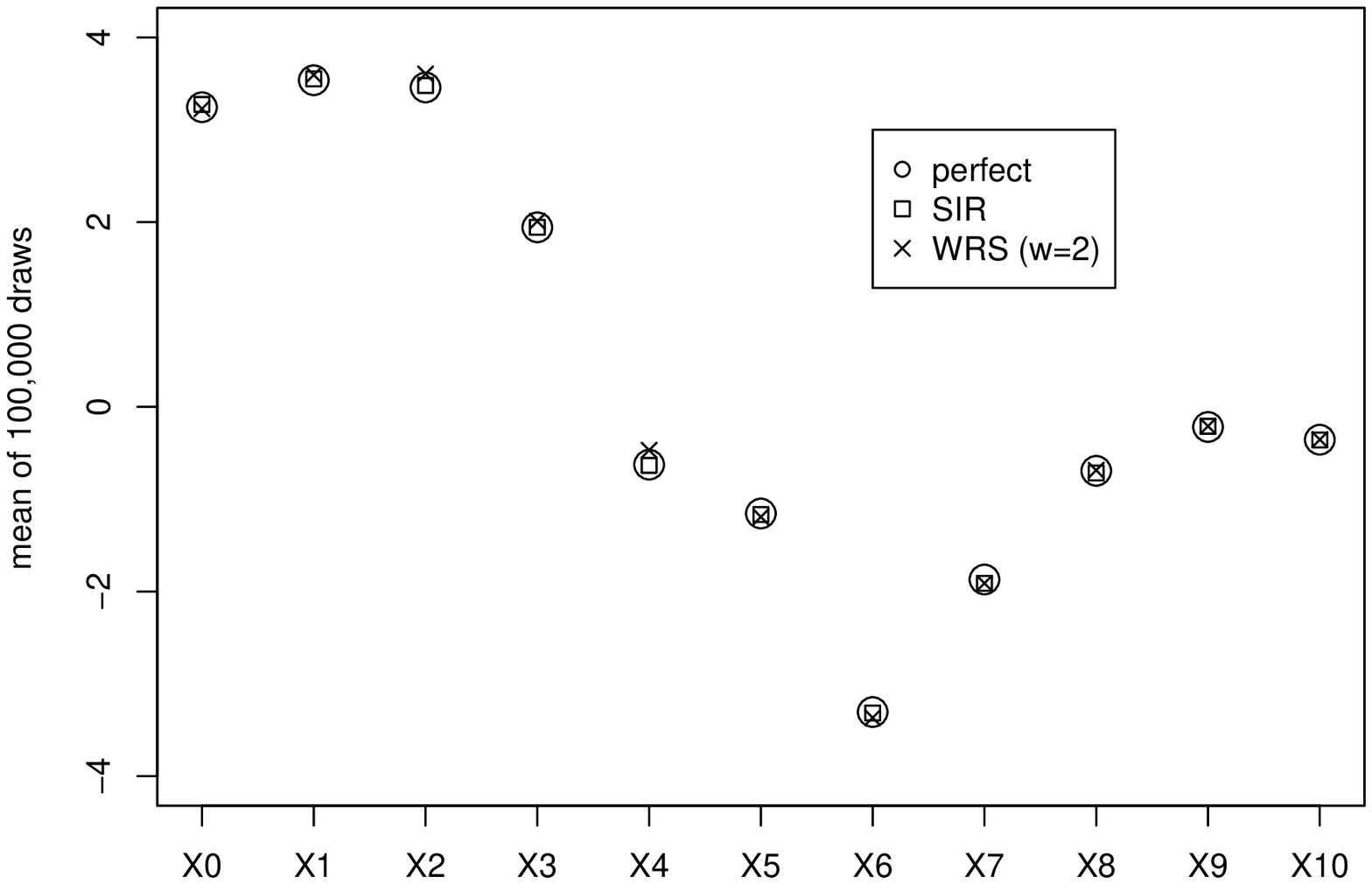}\\
\vspace{-0.1in}
\includegraphics[width=4in,height=2.9in]{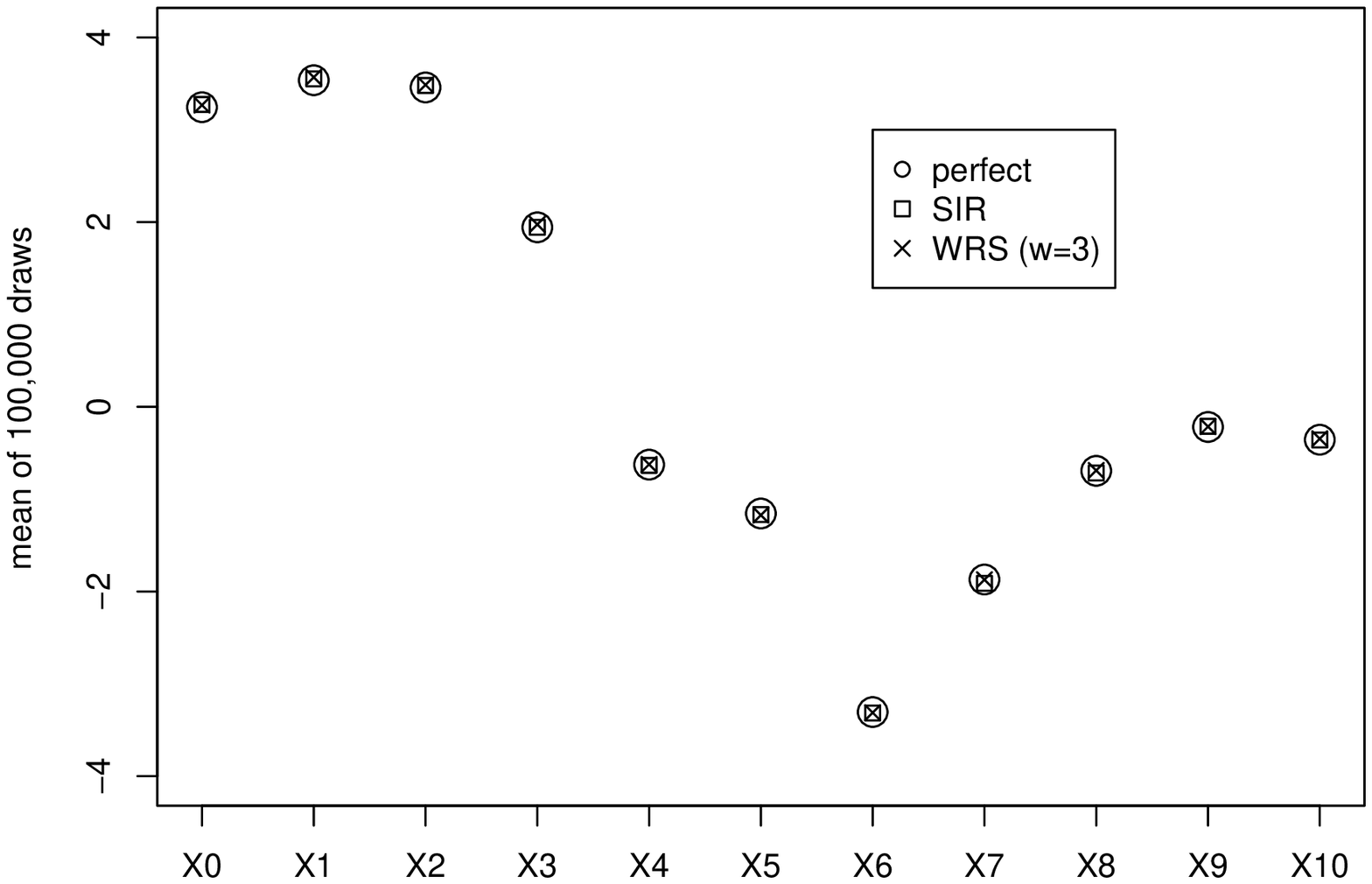}\\
\end{figure}

Continuing in this manner, we generated $N=100,000$ independent values for $X_{0:11}$ for this model using parameters $\mu_{0}=3.0$, $\sigma_{0}=2.0$, $a=0.9$, $b=1.2$, $\sigma_{X}=3.0$, and $\sigma_{Y}=2.3$. (We looked first at such a large $N$ in order to really see the resulting distribution without being concerned about sampling variability.) For comparison, we produced $N$ independent draws directly from the 11-dimensional distribution $\pi(x_{0:10}|y_{1:10})$ using 11-dimensional (non-windowed) rejection sampling. (As expected, the acceptance probabilities are quite small in this case and such direct simulation from the target density is not reasonably achieved for much larger $n$.) Finally, we produced $N$ dependent draws approximately from $\pi(x_{0:10}|y_{1:10})$ using the SIR algorithm. Figure \ref{fig:example1} shows the average marginal values for $X_{0}$ through $X_{10}$ for each algorithm. Only the WRS results change between graphs, showing the anticipated increasing accuracy of the WRS algorithm as the window length increases. For the given model and parameters, a window length of $w=3$ appears to be sufficient. Indeed, as the perfect high-dimensional rejection sampling algorithm is a special case of the WRS algorithm for fixed $n$ and maximal window length, it is easy to code the WRS algorithm once and run it first to get perfect draws from $\pi(x_{0:n}|y_{1:n})$ for the highest feasible $n$ and then to run it with lower $w$, gradually increasing $w$ until sample statistics such as those shown in Figure \ref{fig:example1} reach the desired accuracy.

\begin{figure}[!ht]
  \captionsetup[subfigure]{labelformat=simple}
  \centering
  \caption{Section 4.1 Example: Histogram of values of $X_{7}$ from $100,000$ draws from $\pi(x_{0:11}|y_{1:10})$ using WRS and SIR algorithms} 
  \label{fig:example1a}
  \subfloat[]{\includegraphics[width=.45\textwidth]{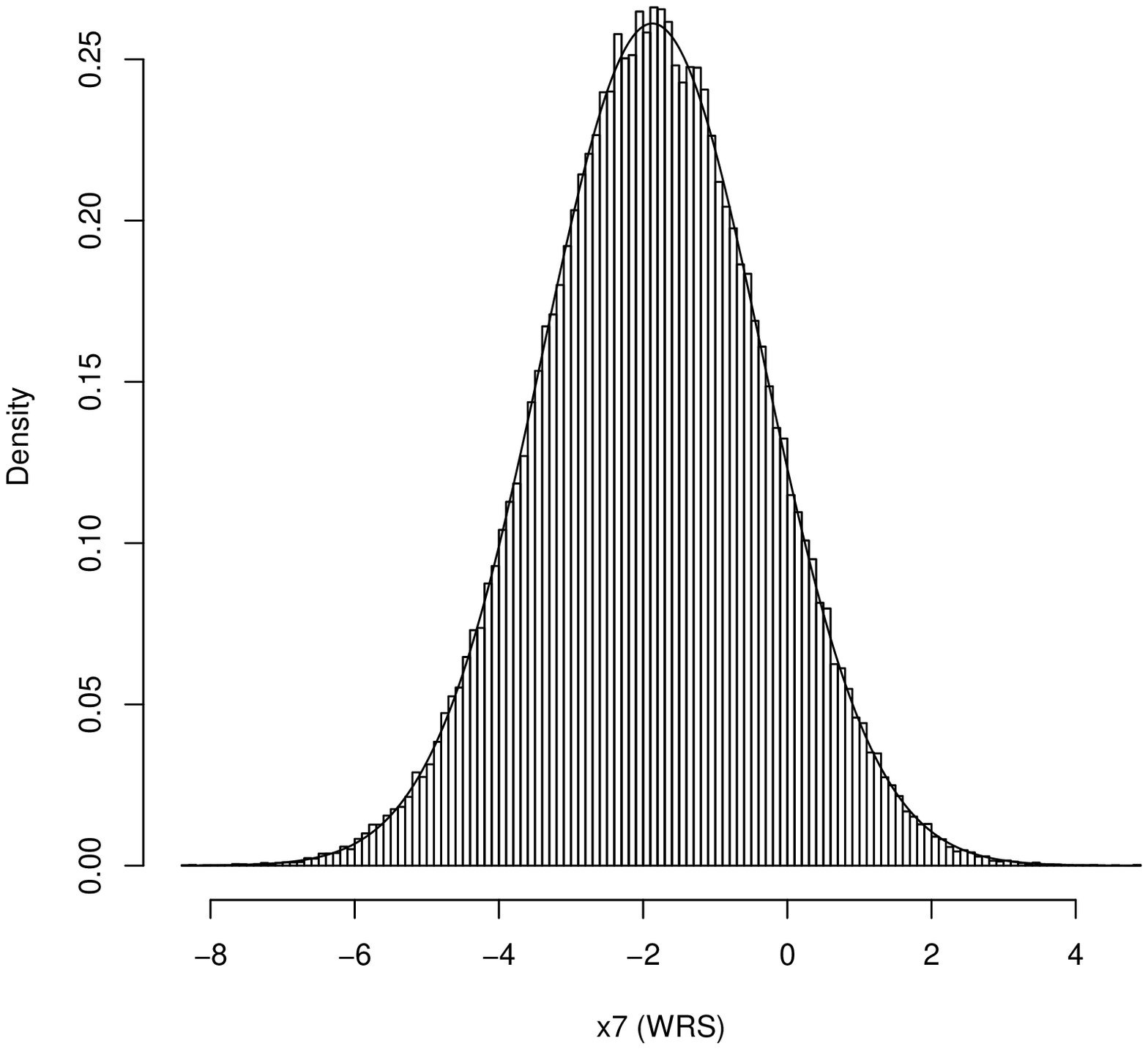}}\qquad
  \subfloat[]{\includegraphics[width=.45\textwidth]{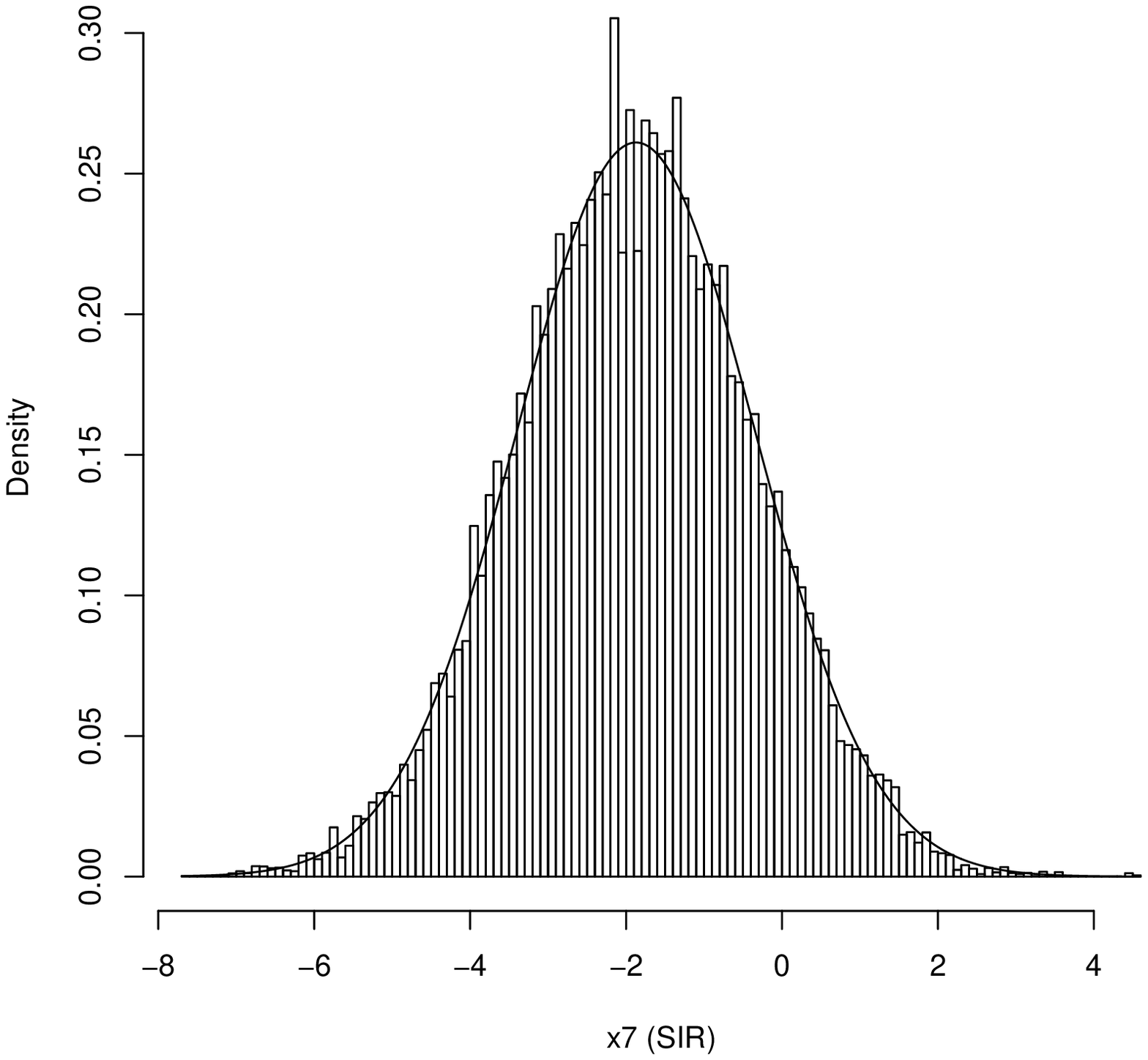}}\\
\vspace{-0.1in}
(a) WRS ($w=3$)\qquad \qquad \qquad \qquad \qquad \qquad \qquad  (b) SIR ($\approx 21.3\%$ distinct)
\end{figure}

For this example, the WRS algorithm produced $100,000$ iid draws from an approximation to $\pi(x_{0:11}|y_{1:10})$ whereas the SIR algorithm (resampling every time) produced dependent draws with only roughly $50\%$ unique 11-dimensional values and only about $10\%$ unique values for $X_{0}$ marginally. Figure \ref{fig:example1a} shows marginal distributions for $X_{7}$ produced by the WRS and SIR algorithms. The overlayed curves are the target normal densities that can be computed analytically for this simple illustrative model. The time to run the WRS and SIR algorithms were comparable. Coded in C++, the WRS algorithm completed in less than $2$ seconds on a standard\footnote{We wish to convey relative speed between algorithms in lieu of particular machine specifications.} laptop while the SIR algorithm completed in 8-10 seconds. The SIR algorithm would obviously speed up if resampling was not done at every step and would likely be the faster algorithm if both were programmed in R where resampling is efficient and accept/reject looping is inefficient.

\begin{figure}[!ht]
  \captionsetup[subfigure]{labelformat=simple}
  \centering
  \caption{Section 4.1 Example: Histogram of values of $X_{1000}$ from $100,000$ draws from $\pi(x_{0:1000}|y_{1:10})$ using WRS and SIR algorithms} 
  \label{fig:example1b}
  \subfloat[]{\includegraphics[width=.45\textwidth]{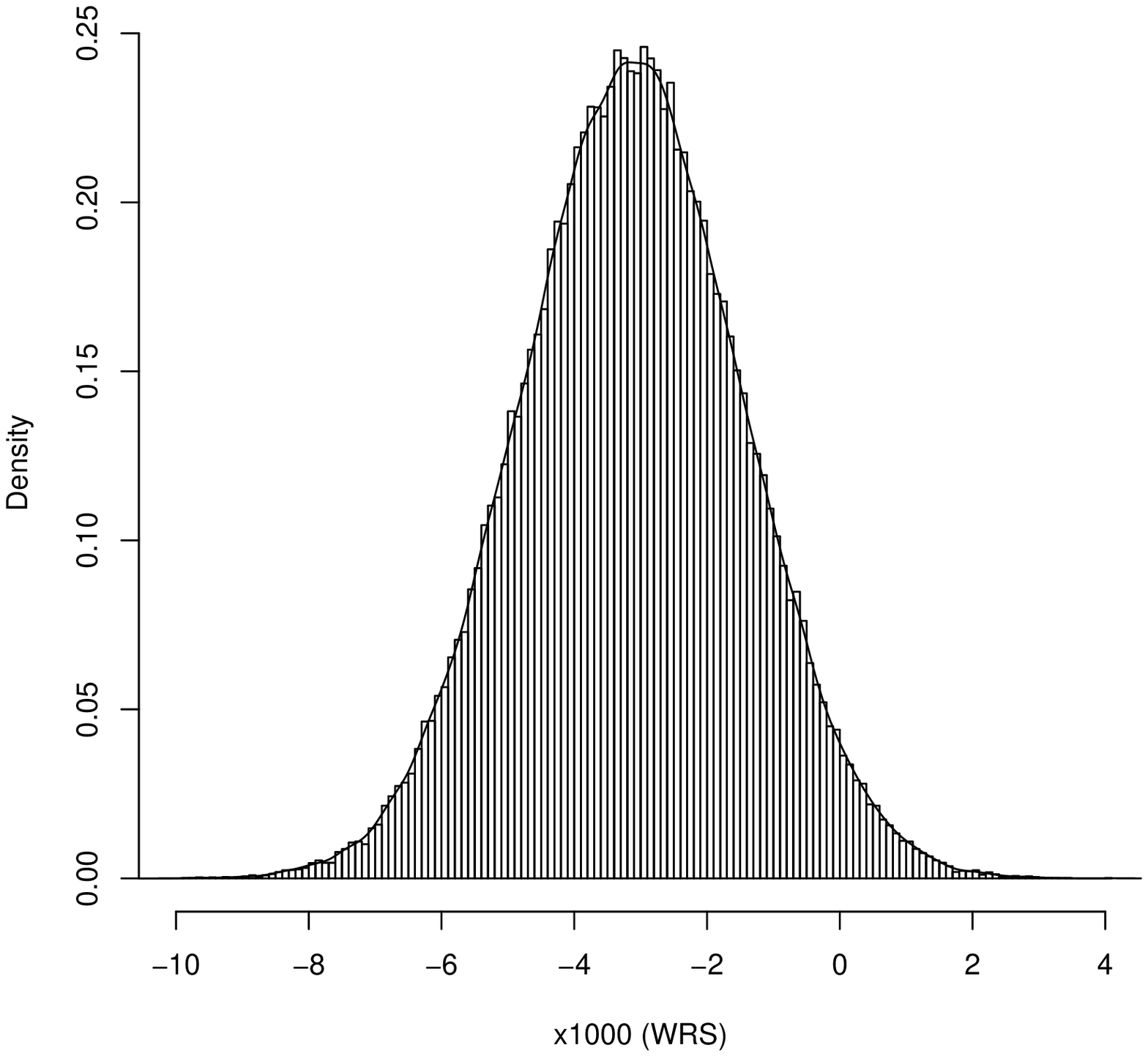}}\qquad
  \subfloat[]{\includegraphics[width=.45\textwidth]{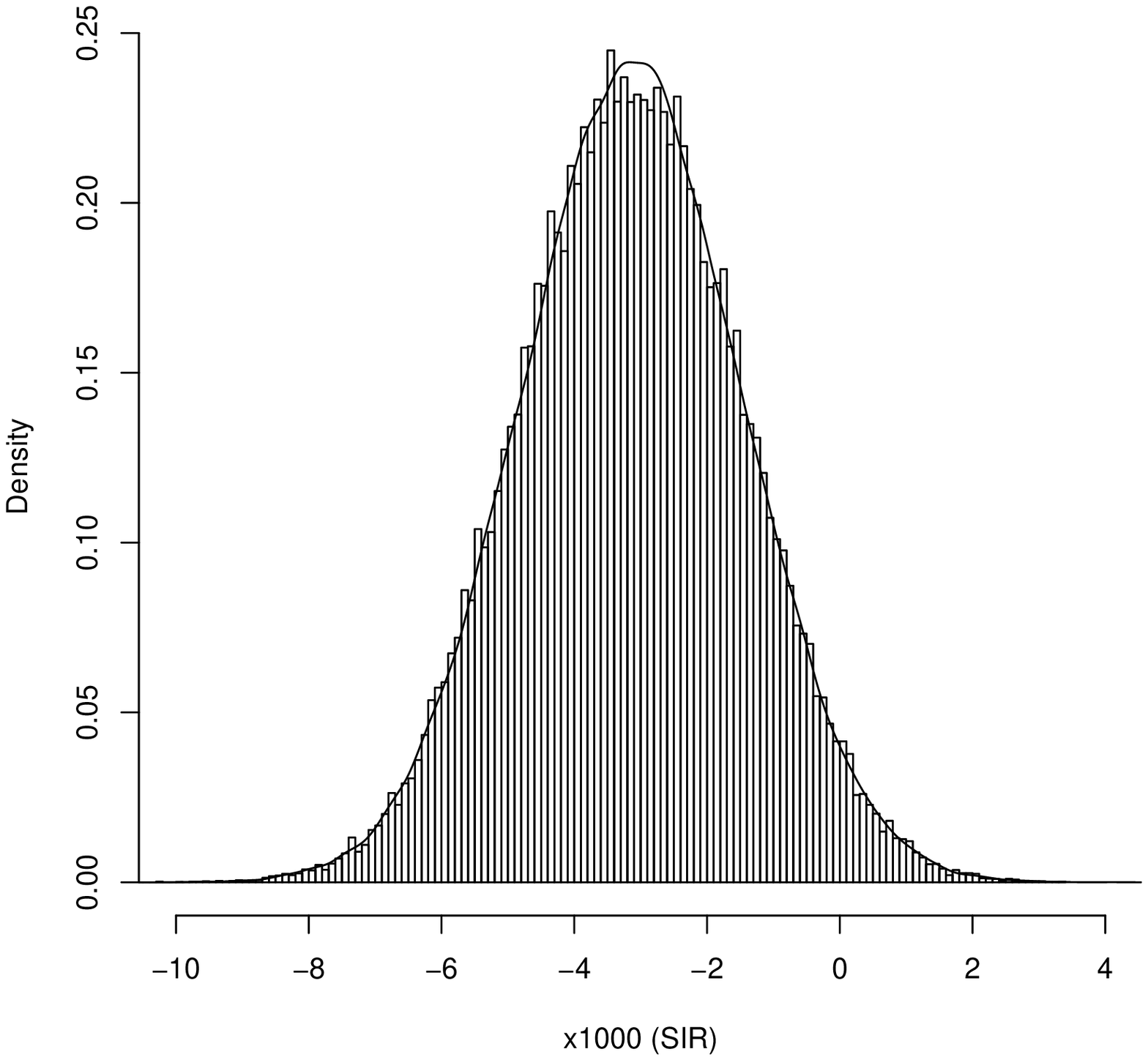}}\\
\vspace{-0.1in}
(a) WRS ($w=3$)\qquad \qquad \qquad \qquad \qquad \qquad \qquad  (b) SIR ($\approx 52\%$ distinct)
\end{figure}

\begin{figure}[!ht]
  \captionsetup[subfigure]{labelformat=simple}
  \centering
  \caption{Section 4.1 Example: Histogram of values of $X_{300}$ from $100,000$ draws from $\pi(x_{0:1000}|y_{1:10})$ using WRS and SIR algorithms} 
  \label{fig:example1c}
  \subfloat[]{\includegraphics[width=.45\textwidth]{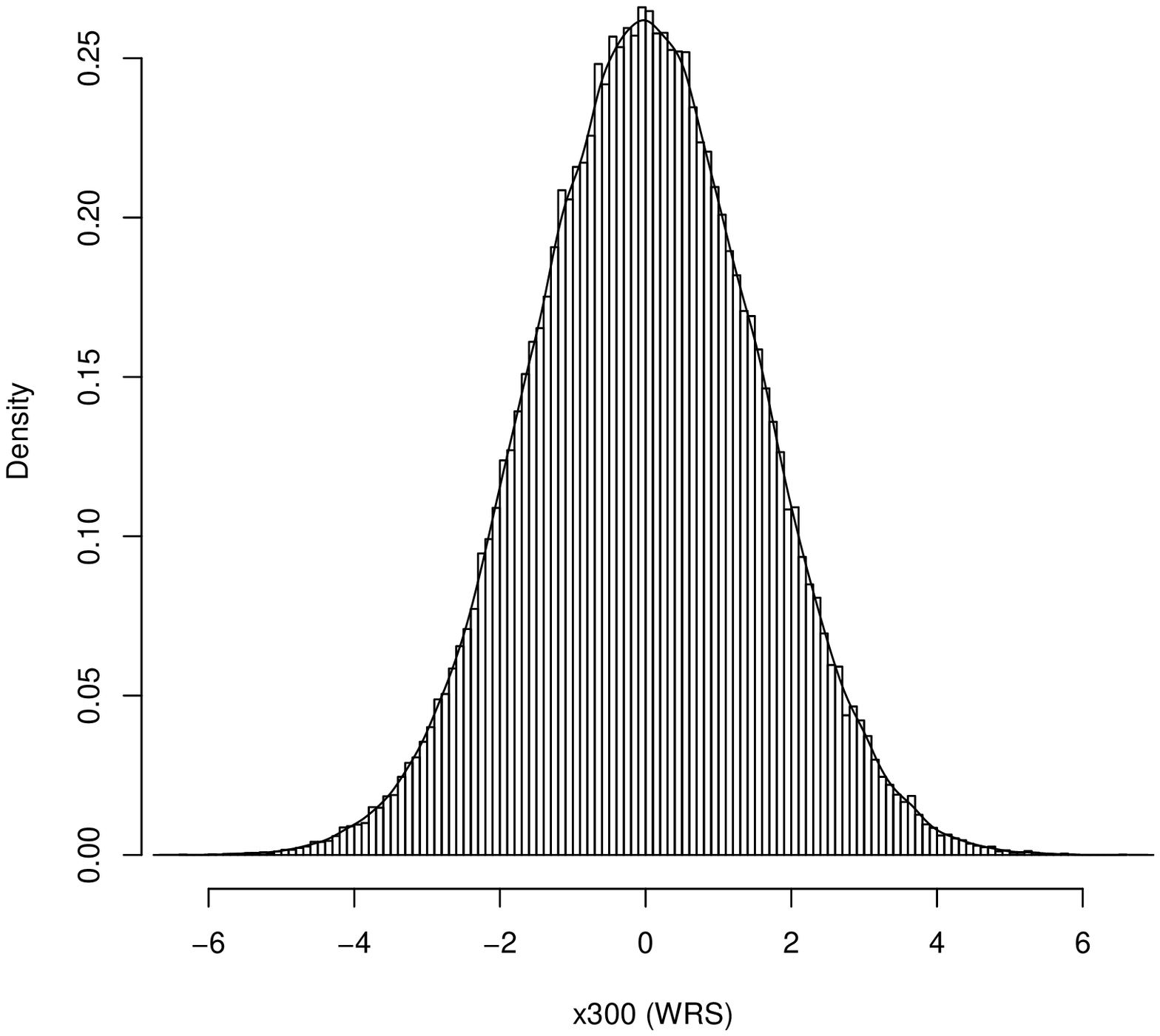}}\qquad
  \subfloat[]{\includegraphics[width=.45\textwidth]{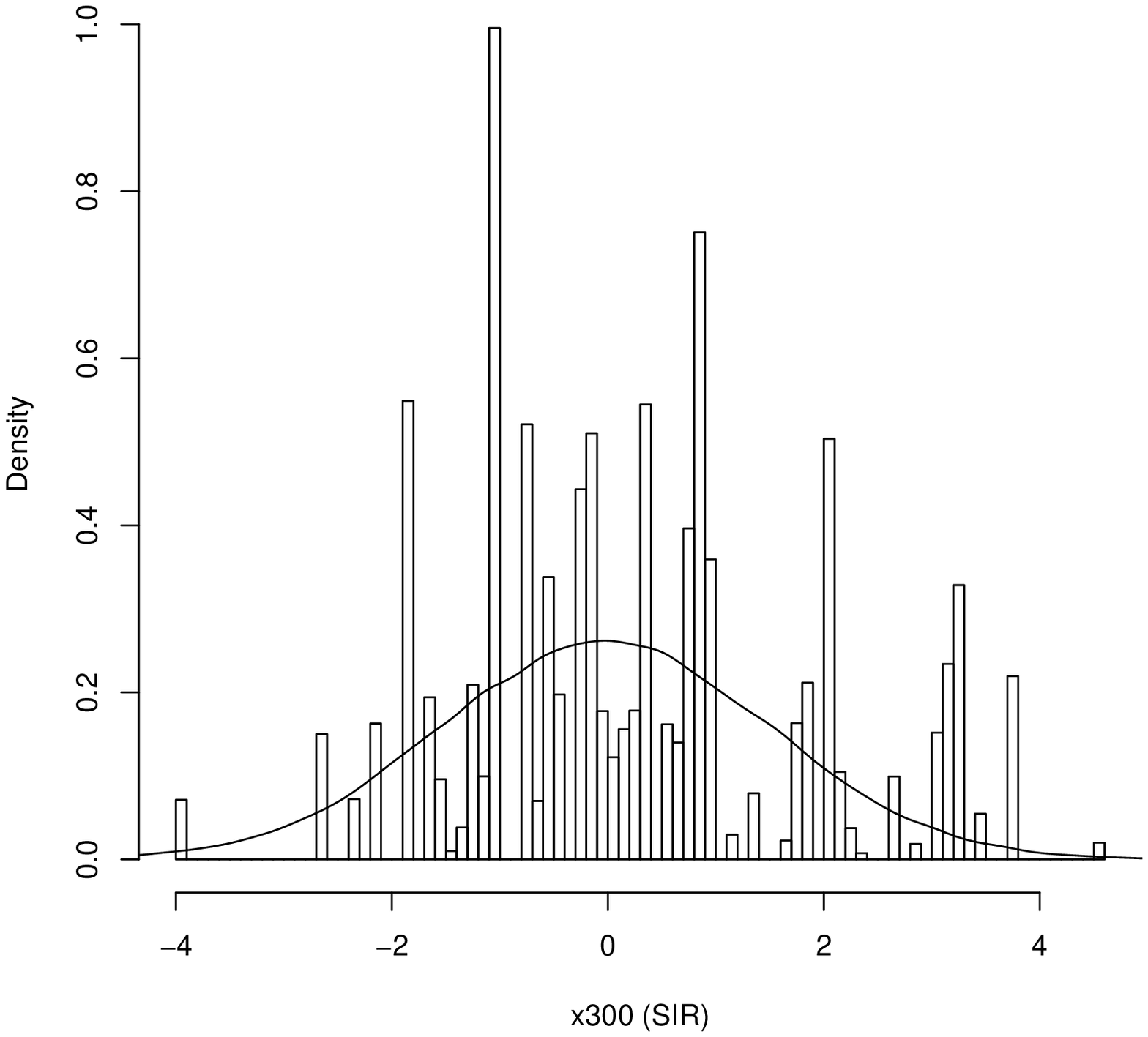}}\\
\vspace{-0.1in}
(a) WRS ($w=3$)\qquad \qquad \qquad \qquad \qquad \qquad \qquad  (b) SIR ($0.07\%$ distinct)
\end{figure}

When comparing the algorithms for larger $n$, it becomes crucial for efficiency to not resample at every time step of the SIR algorithm. As per the discussion at the end of \Section{exist}, we resample only when the effective sample size $N_{eff}$, given by (\ref{e:eff}) falls below the a threshold $N_{thres}$. Following a suggestion in \cite{doucetgod}, we choose $N_{thres}=N/3$. Due to the Markov nature of the model, we are able to continue forward with the window length $w=3$ for increasing $n$. When $n=1,000$, the estimated filtering distribution $\pi(x_{1000}|y_{1:1000})$ is shown in Figure \ref{fig:example1b} for both algorithms. The SIR algorithm resulted in approximately $52\%$ distinct values. An estimated  smoothing distribution, namely an estimate of $\pi(x_{300}|y_{1:1000})$ is shown in Figure \ref{fig:example1c}. It is well known (see \cite{doucet2008}) that the SIR algorithm will produce severely impoverished samples for $\pi(x_{i}|y_{1:n})$ and $i$ small relative to $n$, and in this example we are left with only $0.07$\% distinct values. Suggested approaches to mitigate degeneracy in SIR are discussed in \cite{doucet2008}, but the WRS algorithm performs well without modification. 

%%%%%%%%%%%%%%%%%%%%%%%%%%%%%%%%%%%%%%%%%%%%%%%%%%%%%%%%%%%%%%%
\subsection{Stochastic Volatility Model}
We now consider the {\emph{stochastic volatility}} model, often used in finance and also considered in \cite{doucet2008}, where
$$
X_{0} \sim N (0,\sigma^{2}/(1-\alpha^{2})), \qquad
X_{n+1} = \alpha X_{n} + \sigma \varepsilon_{n+1}, \qquad
Y_{n}= \beta e^{X_{n}/2} \nu_{n}
$$
where $\{\varepsilon_{n}\} \stackrel{iid}{\sim} N(0,1)$ and $\{\nu_{n}\} \stackrel{iid}{\sim} N(0,1)$ are independent sequences. Again we assume that only the $Y_{n}$ are observed.

Following the same procedure in \Section{normalex}, using a general window length $w$, we need to define $q(x_{0:w-1})$, $x_{i}^{*}$ and $q(x_{j:(j+w-1)}|x_{j-1})$. It is easy to show that 
$$
x_{i}^{*} = \arg\max_{x_{i}} \pi(y_{i}|x_{i}) = \arg\max_{x_{i}} N(y_{i};0,\beta^{2} e^{x_{i}})= \ln (y_{i}^{2}/\beta^{2}) \qquad \mbox{for} \,\,\, i=1,2,\ldots.
$$
Taking
$$
q(x_{0}) = \pi(x_{0})=N(x_{0};0,\sigma^{2}/(1-\alpha^{2})) \,\,\, \mbox{and} \,\,\, q(x_{i}|x_{i-1}) = \pi(x_{i|x_{i-1}})=N(x_{i};\alpha x_{i-1},\sigma^{2}),
$$
we have
\beq 
\elabel{firstq}
q(x_{0:w-1}) = q(x_{0}) \prod_{i=1}^{w-1} q(x_{i}|x_{i-1}) 
\eeq
and
\beq
\elabel{secondq}
q(x_{j:(j+w-1)}|x_{j-1}) = \prod_{i=j}^{j+w-1} q(x_{i}|x_{i-1}).
\eeq

\begin{figure}[!ht]
\begin{center}
  \caption{Section 4.2 Example: Means for marginal components in $100,000$ draws from $\pi(x_{0:10}|y_{1:10})$ using perfect 11-dimensional rejection sampling, the SIR algorithm, and the WRS algorithm with $w=5$.}
  \label{fig:example2}
  \includegraphics[height=4in,width=5in]{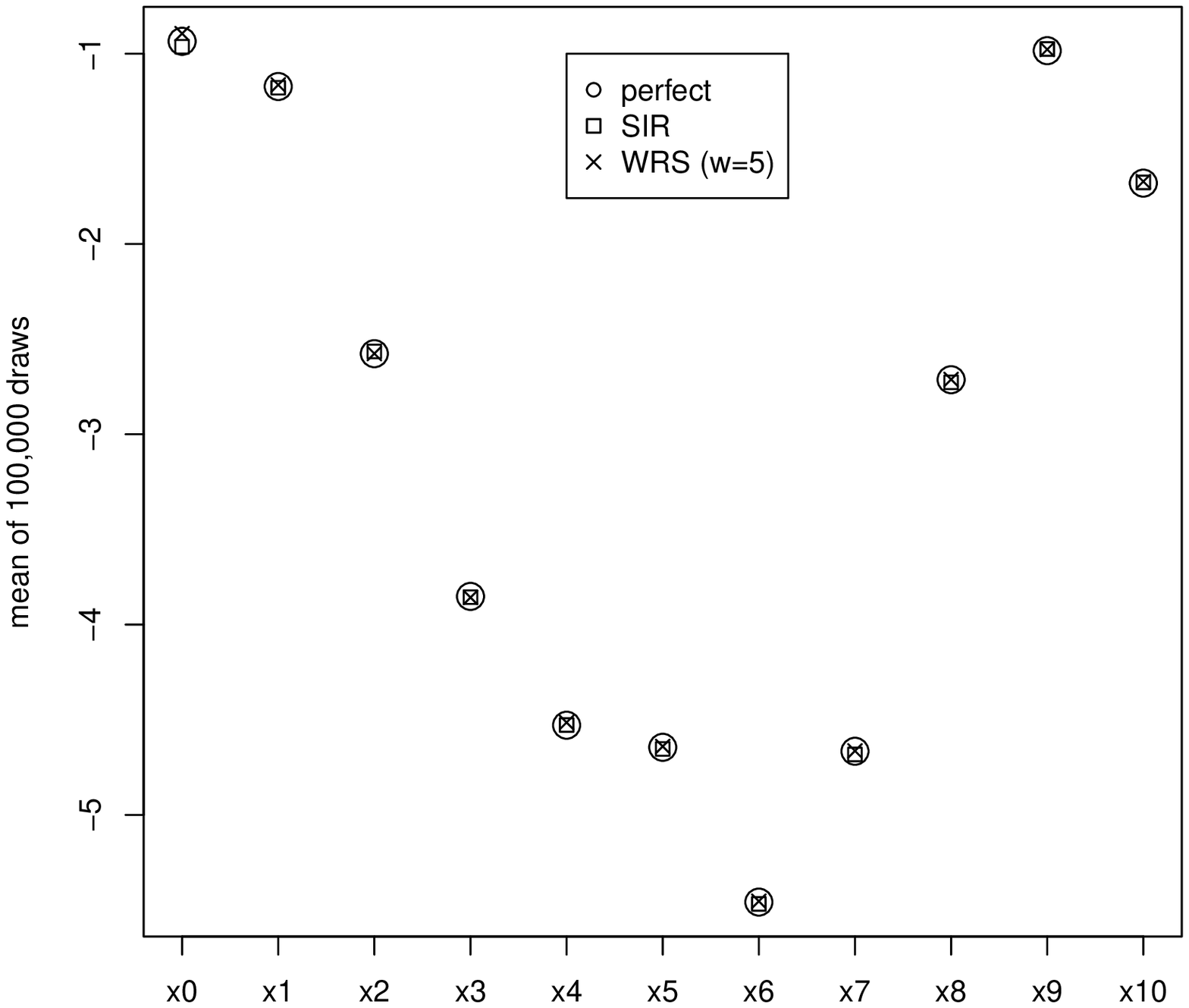}
\end{center}
\end{figure}

\begin{figure}[!ht]
\begin{center}
  \caption{Section 4.2 Example: Proportion of surviving distinct marginal values using SIR}
  \label{fig:example2b}
  \includegraphics[height=4in,width=5in]{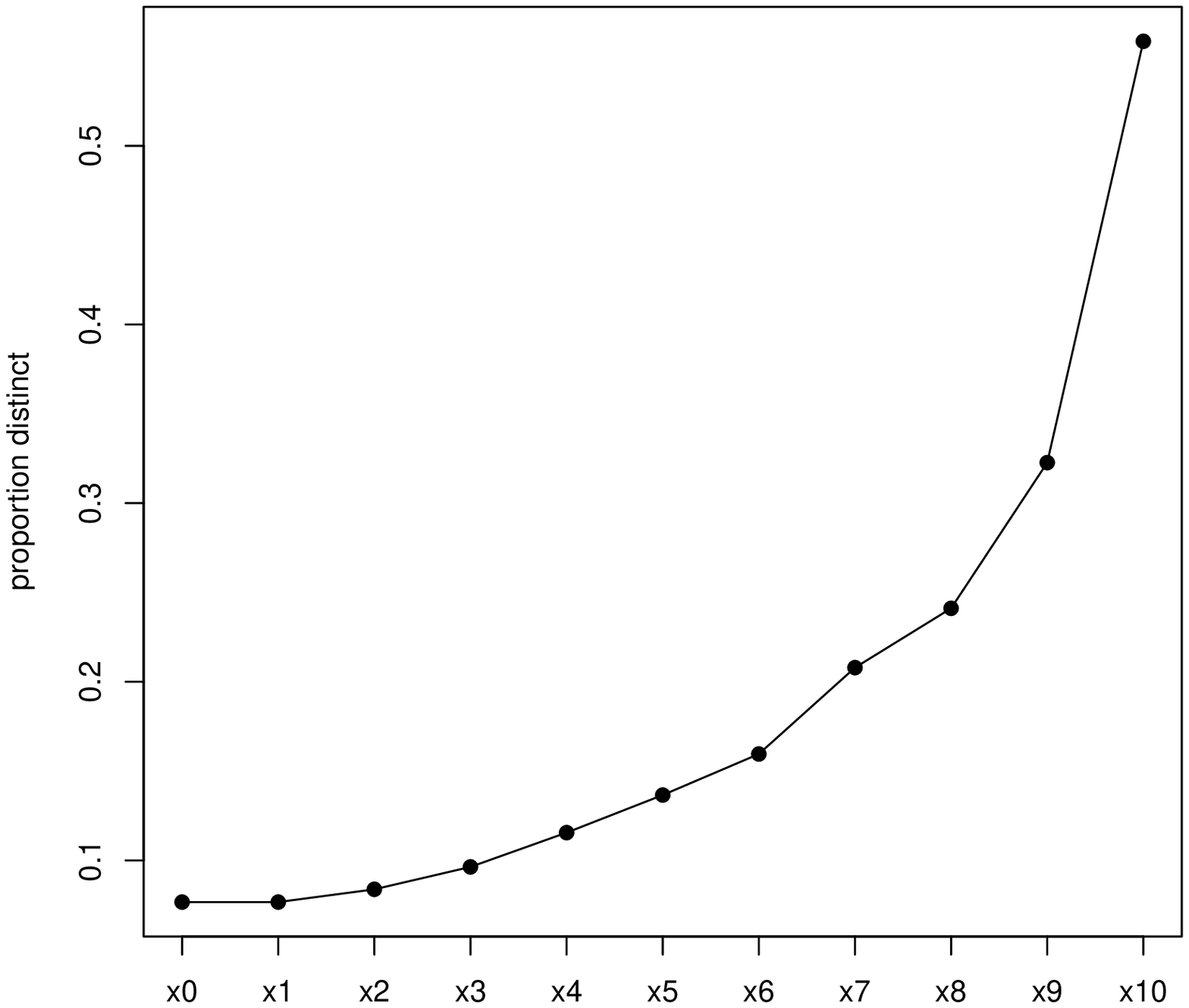}
\end{center}
\end{figure}

\begin{figure}[!ht]
  \captionsetup[subfigure]{labelformat=simple}
  \centering
  \caption{Section 4.2 Example: 100,000 draws from $X_{5}|y_{0:10}$ using WRS and SIR algorithms} 
  \label{fig:example2c}
  \subfloat[]{\includegraphics[width=.45\textwidth]{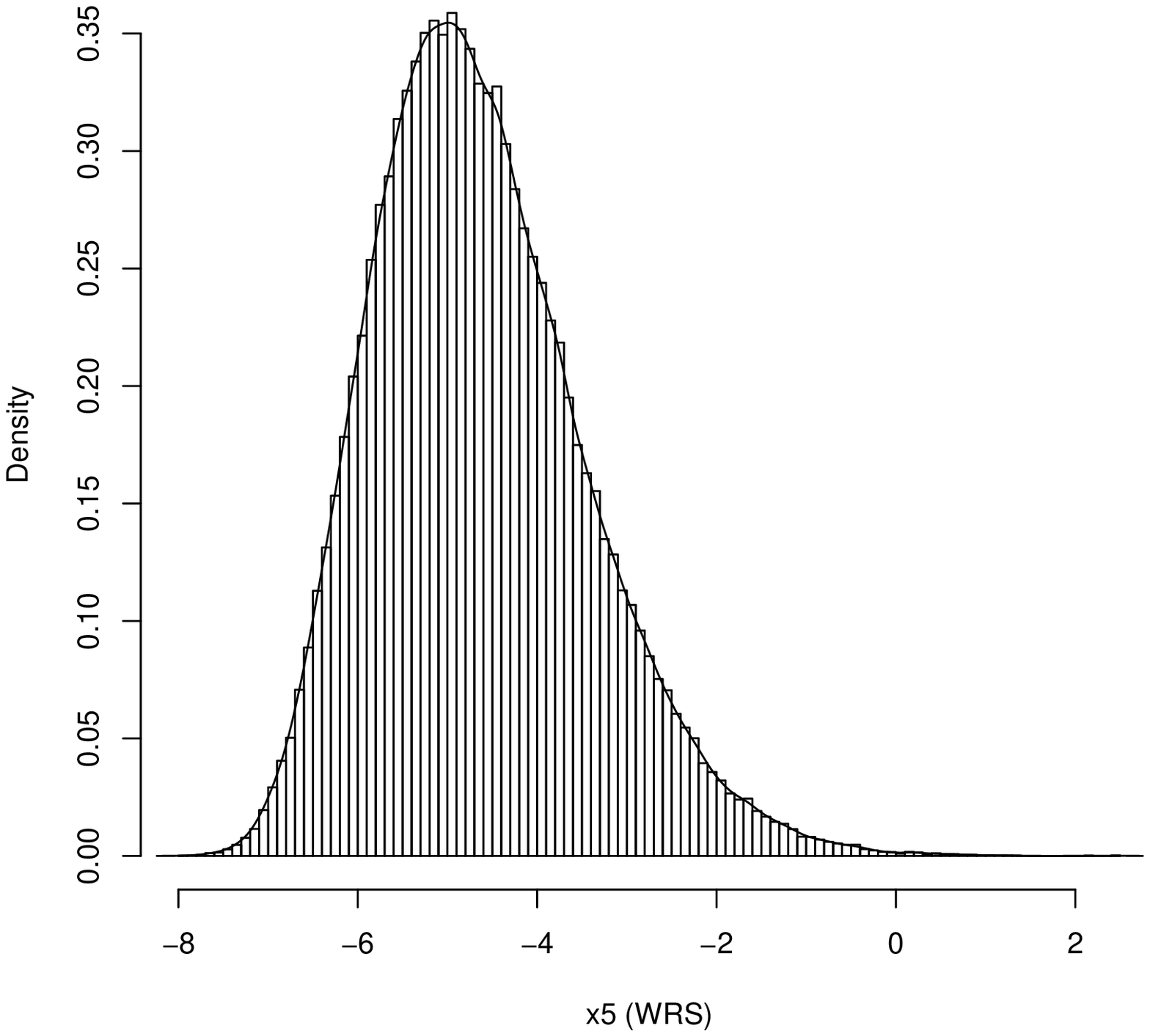}}\qquad
  \subfloat[]{\includegraphics[width=.45\textwidth]{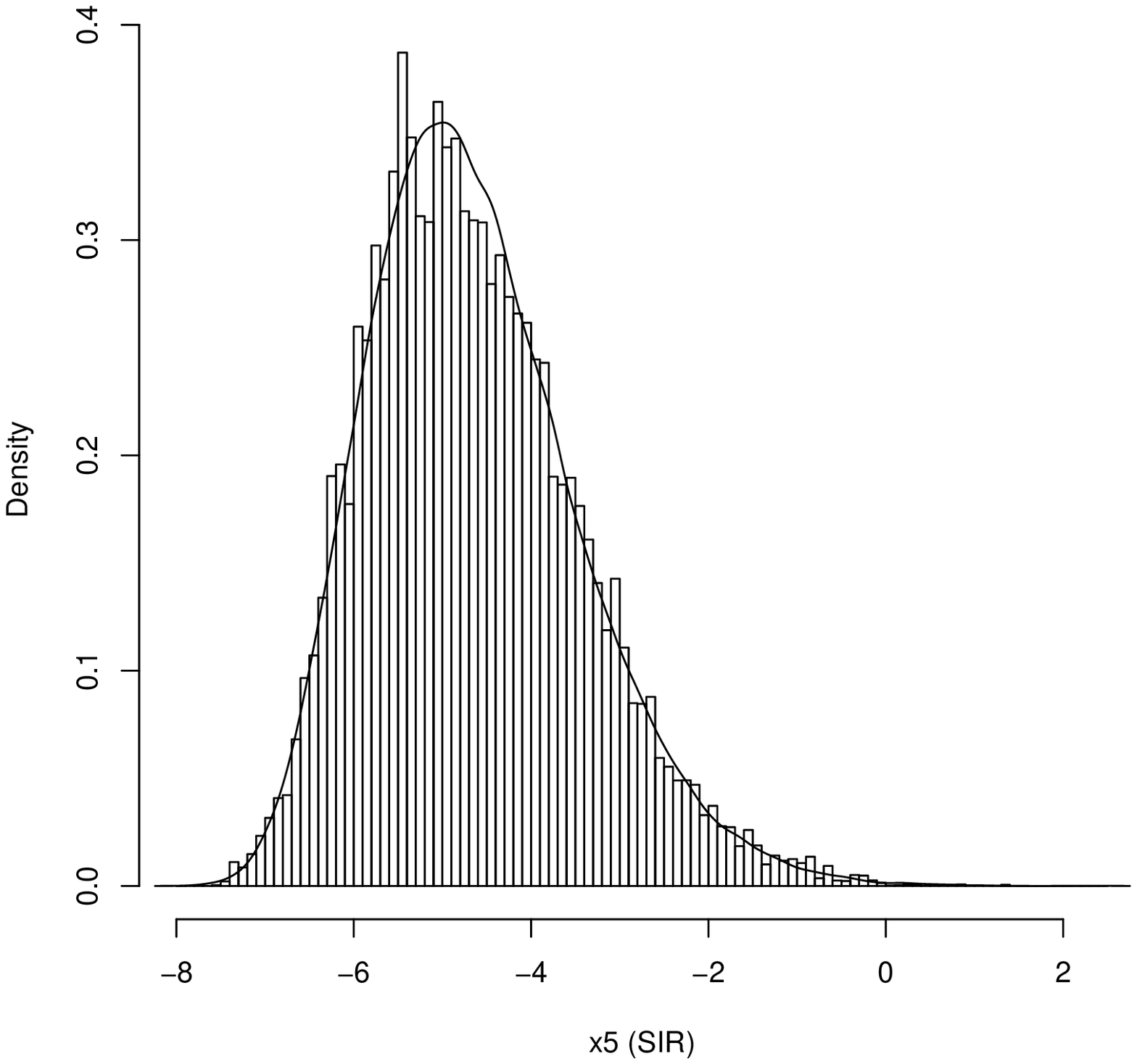}}\\
\vspace{-0.1in}
(a) WRS ($w=5$)\qquad \qquad \qquad \qquad \qquad \qquad \qquad  (b) SIR ($13.6\%$ distinct)
\end{figure}

We ran Step 1 of the WRS algorithm (Algorithm \ref{wrsalg}) with $w=11$ in order to get $N=100,000$ perfect draws via rejection sampling from $\pi(x_{0:10}|y_{1:10})$. ($n=10$ was chosen as the maximal value after which rejection sampling from $\pi(x_{0:n}|y_{1:n})$ became prohibitively slow.) We then ran the complete WRS algorithm with $n=10$ and increasing $w$, starting from $w=1$, until the marginal sample means from WRS for $X_{i}|y_{1:10}$ produced rough (non-simultaneous) two standard deviation confidence intervals which contained the perfect draw estimates of the means. That is, if $\hat{\mu}_{i}$ denotes the sample mean of the $100,000$ values of $X_{i}$ resulting from rejection sampling from $\pi(x_{0:10}|y_{1:10})$ and $\overline{X}_{i}$ and $S_{i}^{2}$ denote the sample means and variances of the marginal distributions produced by WRS, $w$ was increased until $\hat{\mu}_{i} \in \overline{X}_{i} \pm 2S_{i}/\sqrt{N}$ for all $i \in \{0,1,\ldots, 10\}$. This resulted in a window length of $5$. Figure \ref{fig:example2} shows the sample means for the marginal draws with WRS and $w=5$ aligning with those from both perfect and SIR draws. Figure \ref{fig:example2b} shows the proportion of surviving distinct values for each marginal distribution using SIR, which is in contrast to the WRS algorithm giving 100\% distinct values for all marginals. Figure \ref{fig:example2c} shows histograms of the marginal distribution of $X_{5}|y_{1:10}$ obtained from the draws from $\pi(x_{0:10}|y_{1:10})$ using 11-dimensional rejection sampling, the SIR algorithm, and the WRS algorithm with $w=5$. The overlayed curve in each case is the density estimated by 100,000 perfect draws of $X_{5}$ marginalized from 11-dimensional rejection sampling. The SIR algorithm results stand out as the roughest estimate, though, again, it is well known \cite{doucet2008} that this ``straight'' application of the SIR algorithm does not produce good smoothing estimates and that additional steps should be taken to improve it. The WRS algorithm, on the other hand gives a nice estimate of the smoothing distribution straight away with a speed that is comparable to the initial pass of the SIR algorithm. (By ``comparable" we mean that, for the examples in this paper as well as others we tried, the WRS algorithm was  either faster than the SIR algorithm or slower but still finishing only a few seconds behind. In the worst case, the WRS algorithm was about one minute behind the SIR algorithm. No particular effort was made to fully optimize either code.)

%\begin{figure}[!ht]
%  \captionsetup[subfigure]{labelformat=simple}
%\begin{center}
%  \caption{Section 4.2 Example: 100,000 draws from $X_{5}|y_{0:10}$, (a) perfectly, (b) using SIR (13.6\% distinct), and (c) WRS ($w=5$)} 
%  \label{fig:example2c}
%  \subfloat[]{\includegraphics[width=.28\textwidth]{Figures/hist_x5.eps}}\qquad
%  \subfloat[]{\includegraphics[width=.28\textwidth]{Figures/hist_x5_sir.eps}}
%\qquad
%  \subfloat[]{\includegraphics[width=.28\textwidth]{Figures/hist_x5_wrs.eps}}\\
%\vspace{-0.2in}
%(a) \,\,\,\,\,\,\,\,\,\,\,\,\,\,\,\,\,\,\,\,\,\,\,\,\,\,\,\,\,\,\,\,\,\,\,\,\,\,\,\,\,\,\,\,\,\,\,\,\,\,\,\,\,\,\,\,\,\,\,\,\,\,\,\,\,\,\,\,\,\,\,\,(b) \,\,\,\,\,\,\,\,\,\,\,\,\,\,\,\,\,\,\,\,\,\,\,\,\,\,\,\,\,\,\,\,\,\,\,\,\,\,\,\,\,\,\,\,\,\,\,\,\,\,\,\,\,\,\,\,\,\,\,\,\,\,\,\, (c)
%\end{center}
%\end{figure}

%%%%%%%%%%%%%%%%%%%%%%%%%%%%%%%%%%%%%%%%%%%%%%%%%%%%%%%%%%%%%%%%%%
\subsection{A ``Highly Nonlinear'' Model}
In this section, we apply the WRS algorithm the ``highly nonlinear'' model considered in several papers including \cite{arulampalam}, \cite{chenbakshi}, \cite{doucetgod}, and \cite{gordon}.
$$
\begin{array}{lcl}
X_{0} &\sim & N(\mu,\sigma^{2})\\
\\
X_{n+1} &=& 0.5X_{n} + \frac{25 X_{n}}{1+X_{n}^{2}} + 8 \cos (1.2n)+ \varepsilon_{n+1}\\
\\
Y_{n} &=& 0.05 X_{n}^{2} + \nu_{n}
\end{array}
$$
for independent $\varepsilon_{n} \stackrel{iid}{\sim} N(0,\sigma_{X}^{2})$ and $\nu_{n} \stackrel{iid}{\sim} N(0,\sigma_{Y}^{2})$. The parameters were chosen as $\mu = 0$, $\sigma^{2}=5$, and $\sigma_{X}^{2}=\sigma_{Y}^{2}=10$.

Taking
$$
q(x_{0}) = \pi(x_{0})=N(x_{0};\mu,\sigma^{2}) \,\,\, \mbox{and} \,\,\, q(x_{i}|x_{i-1}) = \pi(x_{i}|x_{i-1})=N(x_{i};,m(x_{i-1}),\sigma_{X}^{2}),
$$
with $m(x_{i-1}) = 0.5x_{i-1}+25x_{i-1}/(1+x_{i-1}^{2})+8 \cos[1.2(i-1)]$,
we define $q(x_{0:w-1}) $ and $q(x_{j:(j+w-1)}|x_{j-1})$ using (\ref{e:firstq}) and (\ref{e:secondq}).

It is easy to show that
$$
x_{i}^{*} = \arg\max_{x_{i}} \pi(y_{i}|x_{i}) = 
\left\{
\begin{array}{lcl}
0 & ,& \mbox{if} \,\, y_{i}<0\\
\pm \sqrt{y_{i}/0.05} & ,& \mbox{if} \,\, y_{i}\geq 0.
\end{array}
\right.
$$

\begin{figure}[!ht]
\begin{center}
  \caption{Section 4.3 Example: Proportion of Surviving Distinct Marginal Values using SIR}
  \label{fig:example3}
  \includegraphics[height=4in,width=5in]{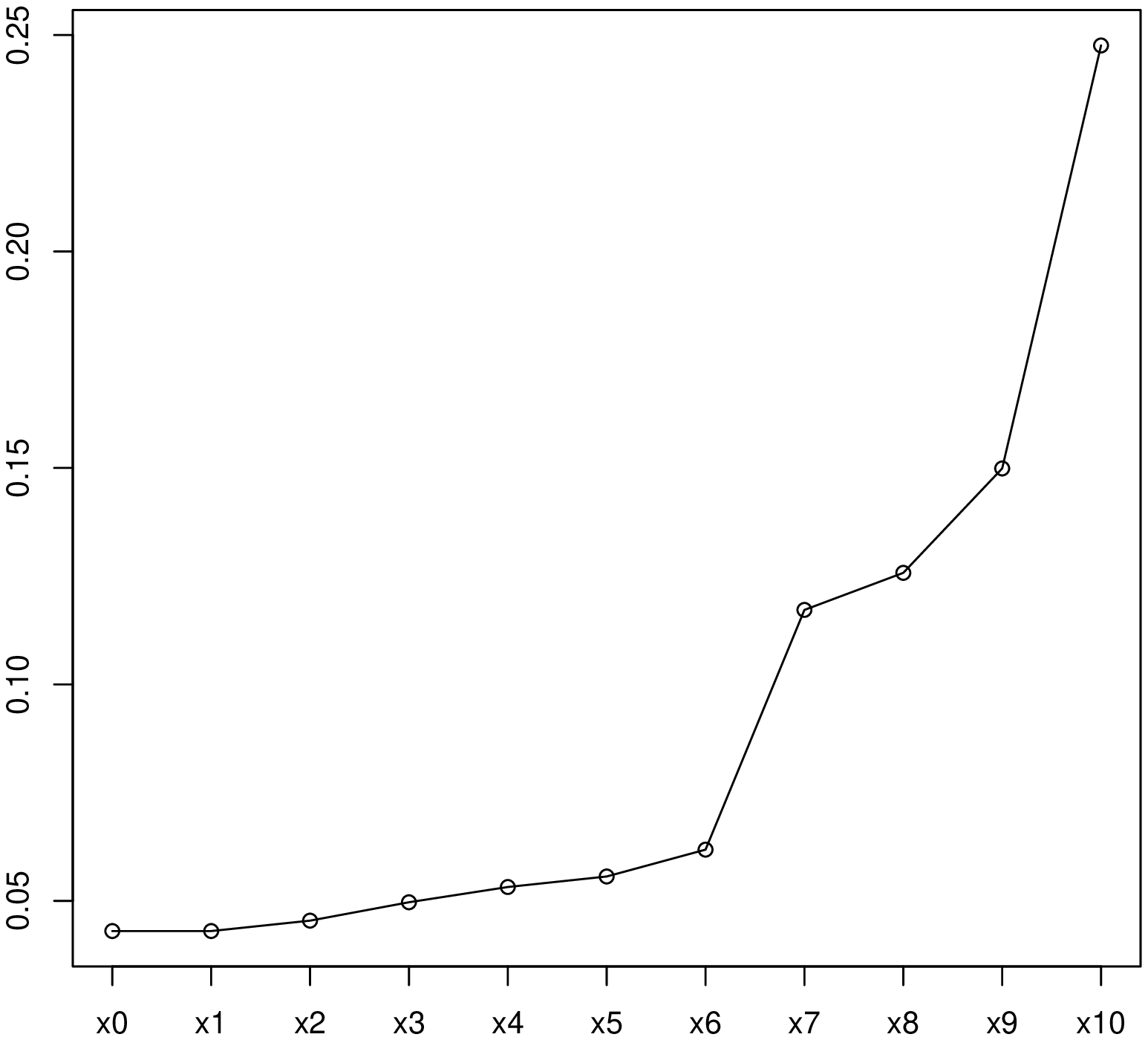}
\end{center}
\end{figure}

\begin{figure}[!ht]
  \captionsetup[subfigure]{labelformat=simple}
  \centering
  \caption{Section 4.3 Example: 100,000 draws from $X_{5}|y_{1:10}$ using WRS and SIR algorithms} 
  \label{fig:example3b}
  \subfloat[]{\includegraphics[width=.45\textwidth]{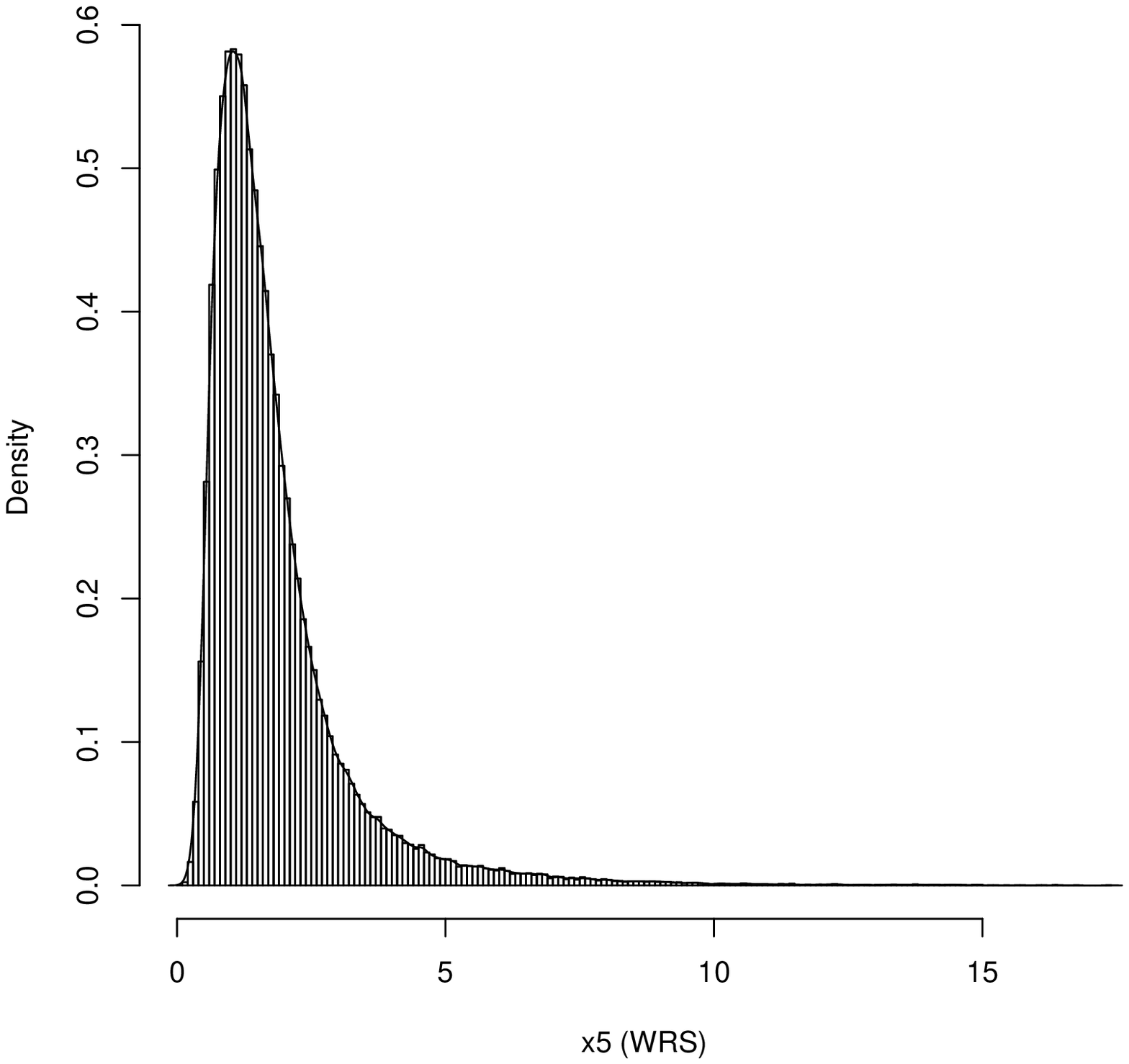}}\qquad
  \subfloat[]{\includegraphics[width=.45\textwidth]{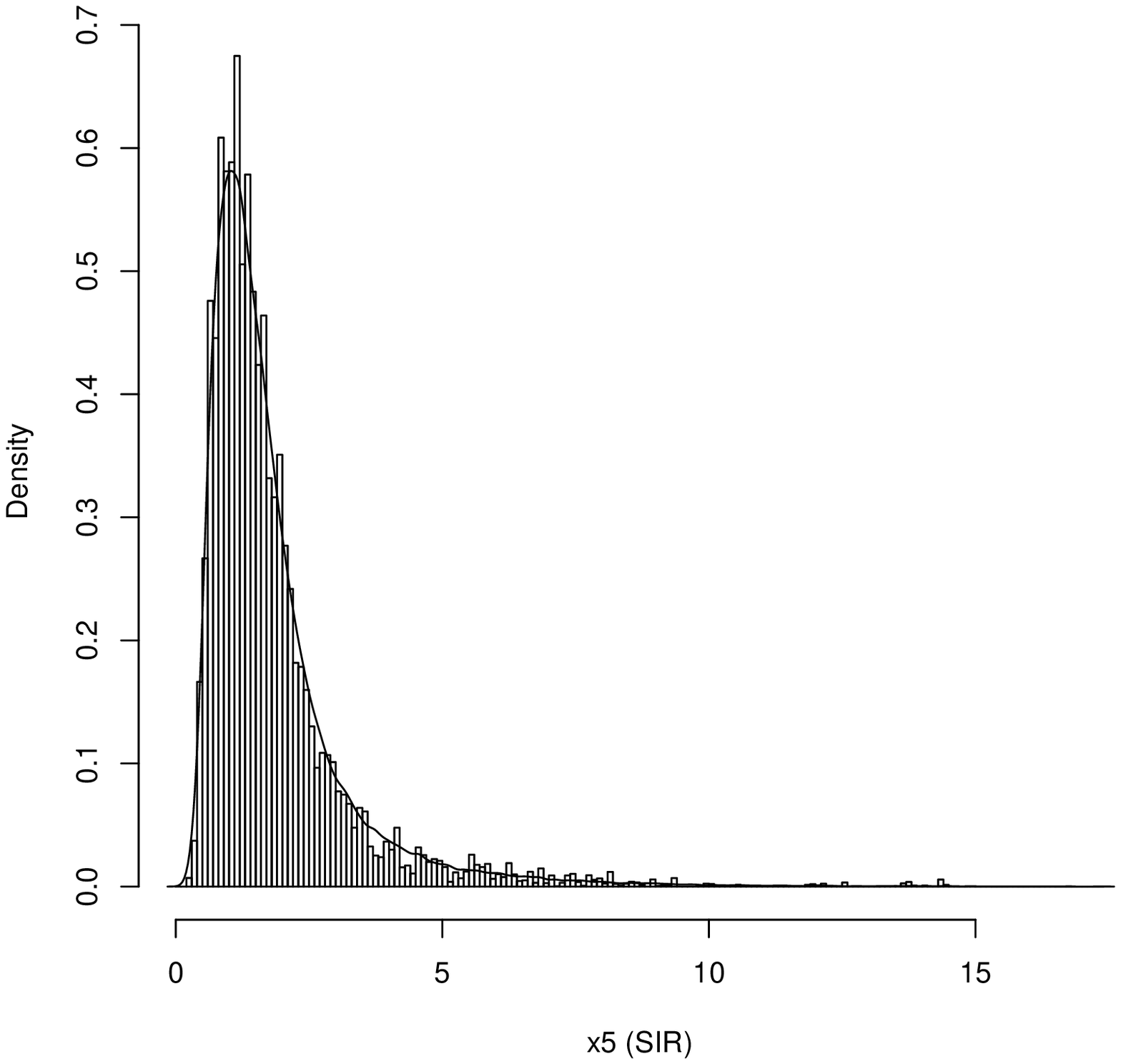}}\\
\vspace{-0.1in}
(a) WRS ($w=4$)\qquad \qquad \qquad \qquad \qquad \qquad \qquad  (b) SIR ($\approx 0.06\%$ distinct)
\end{figure}

An analysis of marginal means such as those depicted in Figures \ref{fig:example1} and \ref{fig:example2} reveal that a window length of $4$ is sufficient for this example. Figure \ref{fig:example3} shows the proportion of surviving distinct values for each marginal distribution using SIR, which, again, is in contrast to the WRS algorithm giving 100\% distinct values for all marginals. Figures \ref{fig:example3b} show histograms of $X_{5}|y_{1:10}$ produced by the WRS and SIR algorithms along with superimposed densities estimated by 11-dimensional rejection sampling. As expected, the WRS results are smoother. In addition, the 100,000 11-dimensional WRS draws are distinct and were produced in about 25 seconds, whereas the SIR results took closer to 1 minute. If the SIR algorithm, again slowed by resampling, had been coded in R, it undoubtedly would have been faster.

%\begin{figure}[!ht]
%  \captionsetup[subfigure]{labelformat=simple}
%\begin{center}
%  \caption{Section 4.3 Example: 100,000 draws from $X_{5}|y_{1:10}$, (a) perfectly, (b) using SIR (13.6\% distinct), and (c) WRS ($w=4$)} 
%  \label{fig:example3b}
%  \subfloat[]{\includegraphics[width=.4\textwidth]{Figures/ex3_x5_perfect.eps}}\qquad
%  \subfloat[]{\includegraphics[width=.4\textwidth]{Figures/ex3_x5_sir.eps}}
%\qquad
%  \subfloat[]{\includegraphics[width=.4\textwidth]{Figures/ex3_x5_wrs.eps}}\\
%\vspace{-0.2in}
%(a) \,\,\,\,\,\,\,\,\,\,\,\,\,\,\,\,\,\,\,\,\,\,\,\,\,\,\,\,\,\,\,\,\,\,\,\,\,\,\,\,\,\,\,\,\,\,\,\,\,\,\,\,\,\,\,\,\,\,\,\,\,\,\,\,\,\,\,\,\,\,\,\,(b) \,\,\,\,\,\,\,\,\,\,\,\,\,\,\,\,\,\,\,\,\,\,\,\,\,\,\,\,\,\,\,\,\,\,\,\,\,\,\,\,\,\,\,\,\,\,\,\,\,\,\,\,\,\,\,\,\,\,\,\,\,\,\,\, (c)
%\end{center}
%\end{figure}

%%%%%%%%%%%%%%%%%%%%%%%%%%%%%%%%%%%%%%%%%%%%%%%%%%%%%%%%%%%%%%%%%%%
\subsection{Two Layer Hidden Markov Model}
\slabel{twolayer}
In \cite{anddouc}, the authors discuss a Rao-Blackwellized particle filtering technique, based on the SIR algorithm, for estimating $\pi(x_{0:n}|z_{1:n})$ in a two-layer hidden Markov model as depicted in Figure \ref{fig:hmm}(b). It involves using SIR to first estimate $\pi(y_{1:n}|z_{1:n})$ and then estimating $\pi(x_{0:n}|z_{1:n})$ using the relationship
$$
\pi(x_{0:n}|z_{1:n}) = \int \pi(x_{0:n}|y_{1:n}) \pi(y_{1:n}|z_{1:n})
$$
and draws from $\widehat{\pi}(y_{1:n}|z_{1:n})$. The WRS algorithm can be used here in place of the SIR algorithm.

In this example, however, we consider a particularly simple two-layer model from \cite{anddouc} which can be dealt with more directly. The ``dynamic tobit model'' is described as follows.
$$
X_{0}  \sim  N \left( 0,\frac{\sigma_{X}^{2}}{1-\phi^{2}} \right)\\
$$
$$
X_{n+1} = \phi X_{n} + \sigma_{X} \varepsilon_{n+1}, \qquad
Y_{n} = X_{n} + \sigma_{Y} \nu_{n}, \qquad
Z_{n} = \max (0,Y_{n})
$$
where $\{\varepsilon_{n}\} \stackrel{iid}{\sim} N(0,1)$ and $\{\nu_{n}\} \stackrel{iid}{\sim} N(0,1)$ are independent sequences. Assume that only the $Z_{n}$ are observed. The parameters were chosen as $\sigma_{X}^{2} = 0.05$, $\sigma_{Y}^{2}=0.30$, and $\phi=0.99$, as in \cite{anddouc}.

Consider the case first when $n=1$. If $z_{1}>0$, we wish to draw from
$$
\pi(x_{0},x_{1}|z_{1}) = \pi(x_{0},x_{1}|y_{1}) \propto h(x_{0},x_{1}|y_{1}) = \pi(y_{1}|x_{1}) \pi(x_{1}|x_{0}) \pi(x_{0})
$$
where $y_{1}=z_{1}$.

To implement the SIR algorithm, we take $q(x_{0},x_{1}) := \pi(x_{1}|x_{0}) \pi(x_{0})$ and use the weight $w(x_{0:1}) := \pi(y_{1}|x_{1})$. To implement rejection sampling, we use the fact that
$$
h(x_{0},x_{1}|y_{1}) \leq M \cdot  q(x_{0},x_{1})
$$
where $M = \left. 1 \middle/ \sqrt{2 \pi \sigma_{Y}^{2}} \right.$ and, again, $q(x_{0},x_{1}) := \pi(x_{1}|x_{0}) \pi(x_{0})$.

If $z_{1}=0$, we wish to draw from 
$$
\pi(x_{0},x_{1}|z_{1}) = \pi(x_{0},x_{1}|y_{1}<0) \propto P_{\pi}(y_{1}<0|x_{1}) \pi(x_{1}|x_{0}) \pi(x_{0})
$$
where $ P_{\pi}(y_{1}<0|x_{1}) = \int_{-\infty}^{0} \pi(y_{1}|x_{1}) \, dy_{1}$. We then have the unnormalized target density
\beq
\elabel{untarget}
h(x_{0},x_{1}|z_{1}) = \Phi (-x_{1}/\sigma_{Y}) \cdot \pi(x_{1}|x_{0}) \pi(x_{0})
\eeq
where $\Phi(\cdot)$ denotes the cumulative distribution function for the standard normal distribution.

Taking $q(x_{0},x_{1}) := \pi(x_{1}|x_{0}) \pi(x_{0})$, we can implement the SIR algorithm to draw from (\ref{e:untarget}) using $w(x_{0:1}) = \Phi (-x_{1}/\sigma_{Y})$ and we can implement rejection sampling using the fact that $\Phi (-x_{1}/\sigma_{Y}) \leq M := 1$.

For general $n$, we use $q(x_{n}|x_{n-1}) = \pi(x_{n}|x_{n-1})$ for both algorithms. The incremental weight for the SIR algorithm is 
$$
\alpha (x_{1:n}) = \left\{
\begin{array}{lcl}
\pi_{Y|X}(z_{n}|x_{n}) &,& \mbox{if}\,\,\, z_{n}>0\\
\\
\Phi(-x_{n}/\sigma_{Y})&,& \mbox{if}\,\,\, z_{n}=0.\\
\end{array}
\right.
$$
Here, $\pi_{Y|X}(\cdot|\cdot)$ is the conditional density for $Y_{n}$ given $X_{n}$, previously identifiable through arguments alone.

For the WRS algorithm with window length $w$, we define $M$ in Step 1 of Algorithm \ref{wrsalg} as
$$
M = \prod_{i=1}^{w} \left\{ \pi_{Y|X}(z_{i}|x_{i}) \ind [z_{i}>0] + \Phi(-x_{i}/\sigma_{Y}) \ind [z_{i}=0] \right\},
$$
and $M$ in Step 2 as
$$
M = \prod_{i=m}^{m+w-1} \left\{ \pi_{Y|X}(z_{i}|x_{i}) \ind [z_{i}>0] + \Phi(-x_{i}/\sigma_{Y}) \ind [z_{i}=0] \right\}.
$$

\begin{figure}[!ht]
\begin{center}
  \caption{Section 4.4 Example: Means for marginal components in $100,000$ draws from $\pi(x_{0:10}|z_{1:10})$ using perfect 11-dimensional rejection sampling, the SIR algorithm, and the WRS algorithm with $w=9$.}
  \label{fig:example4}
  \includegraphics[height=4in,width=5in]{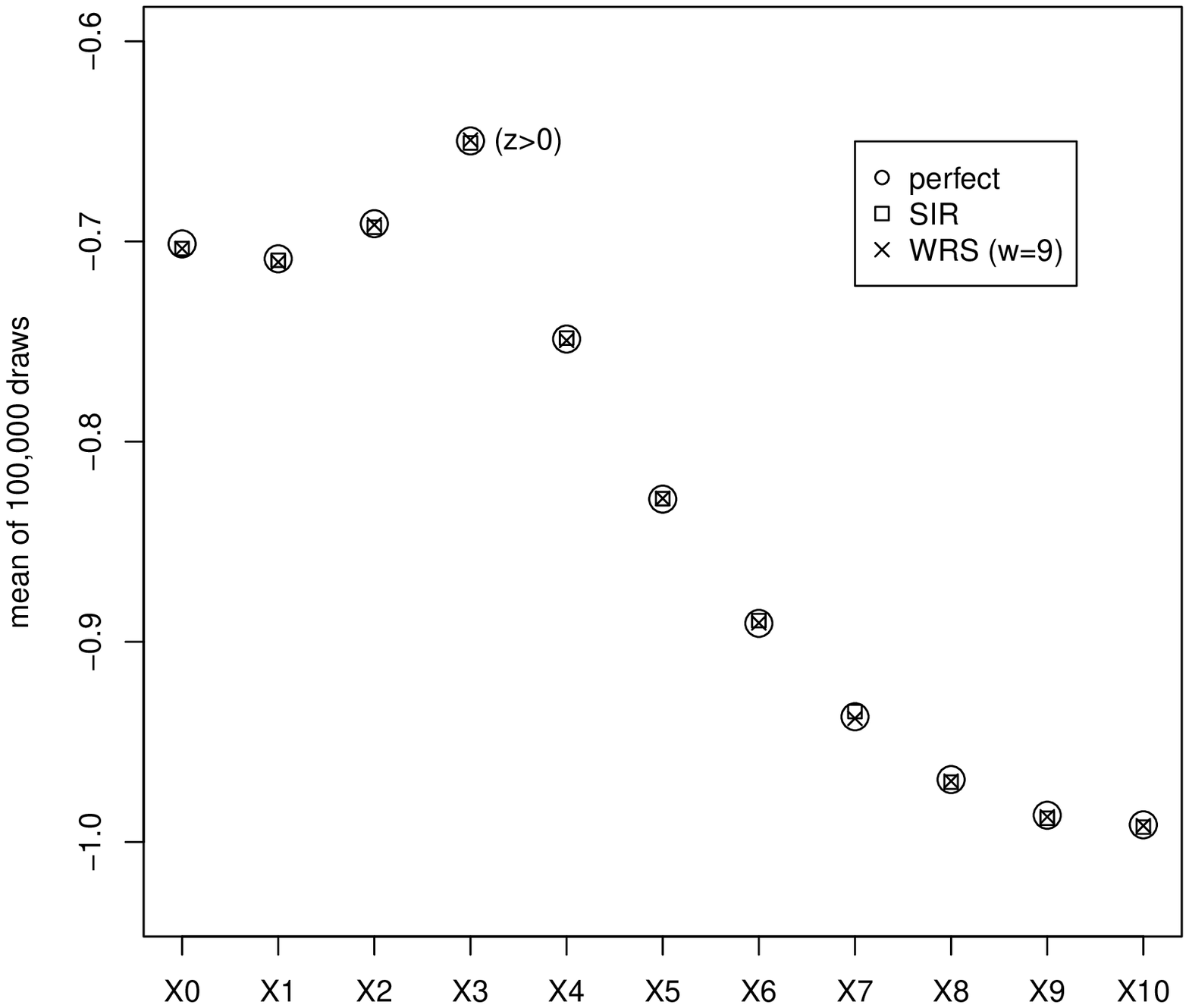}
\end{center}
\end{figure}

Since the underlying hidden Markov model is an AR(1) model with autocorrelation function $\rho(h) = \phi^{h}$, it is not surprising that, for $\phi=0.99$, we would need $w$ to be large for the WRS algorithm.  As in previous examples, we produced $100,000$ independent draws directly from the 11-dimensional distribution $\pi(x_{0:10}|z_{1:10})$ using 11-dimensional, non-windowed, rejection sampling and then used windowed rejection sampling with increasing values of $w$ until we matched marginal means. Figure \ref{fig:example4} shows the matching means, as well as the matching means from the SIR algorithm. The turning point in this graph of means corresponds to the fact that $z_{3}$ was the only data point in $z_{1:10}$ that was positive. Both the WRS and SIR algorithms were quite slow for this example (on the order of 45 minutes) but were still of comparable speeds. 

\begin{figure}[!ht]
\begin{center}
  \caption{Section 4.4 Example: Means for certain marginal components in $100,000$ draws from $\pi(x_{0:1000}|z_{1:1000})$ using the SIR algorithm and the WRS algorithm with $w=9$.}
  \label{fig:example4b}
  \includegraphics[height=4in,width=5in]{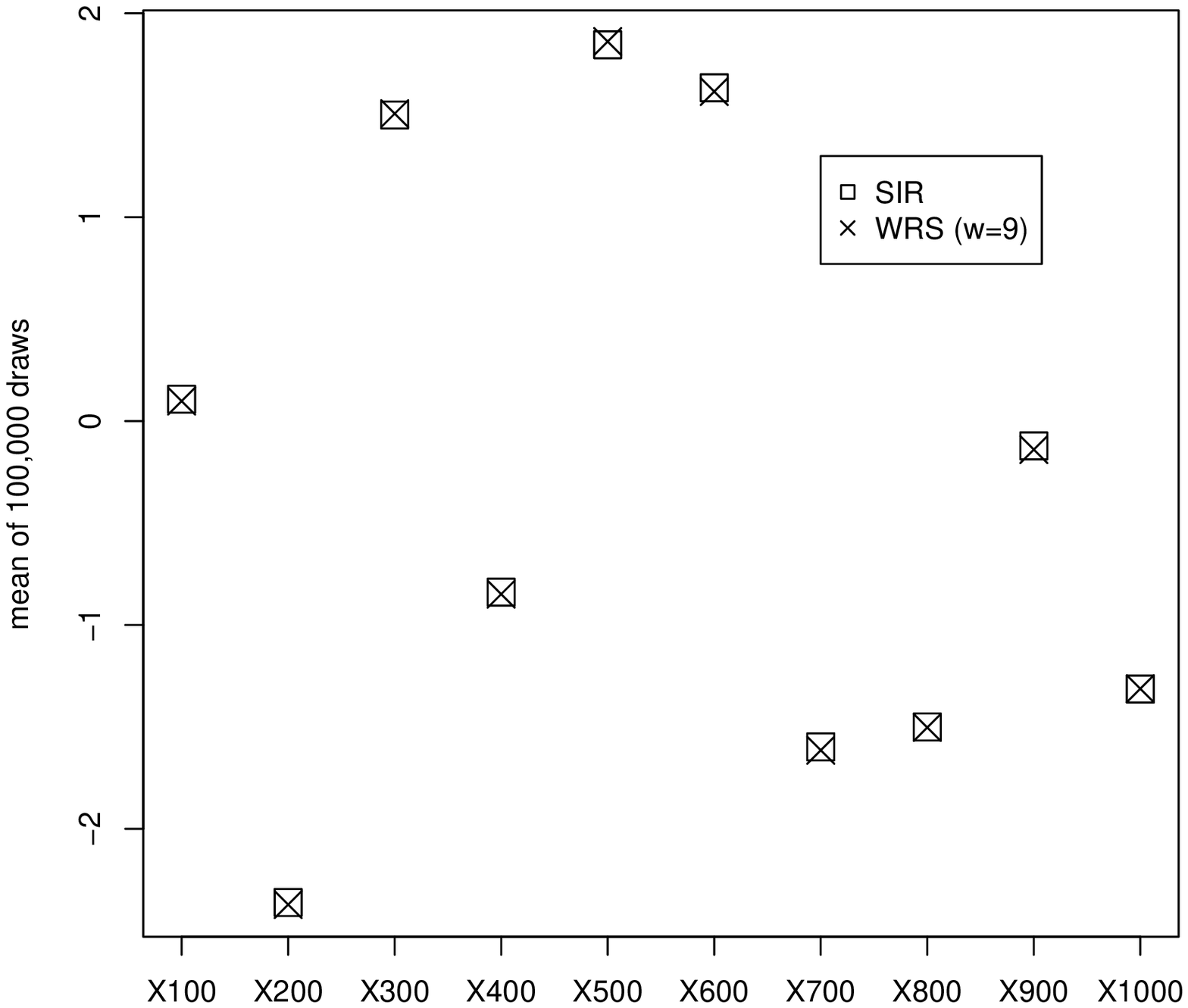}
\end{center}
\end{figure}

In Figure \ref{fig:example4b} we see that the marginal means for the SIR and WRS algorithms continue to coincide even for large $n$. Due to the Markov nature of the model, the window length determined with smaller $n$ can still be used.

%%%%%%%%%%%%%%%%%%%%%%%%%%%%%%%%%%%%%%%%%%%%%%%%%%%%%%%%%%%%%%%%%%%%
%%%%%%%%%%%%%%%%%%%%%%%%%%%%%%%%%%%%%%%%%%%%%%%%%%%%%%%%%%%%%%%%%%%%

\section{Conclusions}

The WRS algorithm produces independent and identically distributed draws from an approximation to $\pi(x_{0:n}|y_{1:n})$ in a semi-sequential manner. That is, for each new observed $y_{n}$, the algorithm produces not only values for $x_{n}$ but reproduces values for some previous $x_{i}$ within a certain window length. At first glance, this redundancy, combined with the use of rejection sampling which has the potential to loop for some time,  makes the WRS algorithm seem impractical. Indeed the WRS algorithm would be very inefficient if it were programmed in R, however in other languages the WRS and SIR algorithms ran at comparable speeds for all of the examples we tried including the benchmark examples from the SIR literature that were included in this paper. Ideally, one would compare speeds between the WRS algorithm programmed in a looping friendly language such as FORTRAN and SIR in a resampling friendly language such as R. When making speed comparisons in this paper, it was not our intentions to ``split hairs'' at this level but only to assert that the WRS can compete and can do so with the added benefit of full samples of distinct iid values.

As for the smoothing problem, our comparison with SIR was also not entirely fair, as it is well known (See, for example, \cite{doucet2008}.) that one should take additional steps such as those involving forward-backward recursions or ``backward information'' filters. However, the point to be made is that the WRS algorithm produces smoothing distributions without any additional effort.

The WRS algorithm is not a good alternative to traditional particle filters for non-Markovian models or in cases where the required minimal window length is so large as to  result in inefficient rejection sampling.

\bibliographystyle{plain}

%\bibliography{C:/BIBS/masterbib}
\bibliography{/home/corcoran/Latex/BIBS/strings,/home/corcoran/Latex/BIBS/masterbib}

\end{document}